\documentclass[preprint,12pt]{elsarticle}

\usepackage{amssymb}
\usepackage{subcaption}
\usepackage{lmodern}
\usepackage{multirow}
\usepackage{hyperref}
\usepackage{amsmath}
\usepackage{setspace}
\usepackage{tabularx}
\usepackage{booktabs} % For better table lines
\usepackage{caption}

\usepackage[linesnumbered, algoruled, vlined]{algorithm2e}

\SetCommentSty{mycommfont}

\usepackage[table,xcdraw]{xcolor}
\usepackage{listings}
\definecolor{mGreen}{rgb}{0,0.6,0}
\definecolor{mGray}{rgb}{0.5,0.5,0.5}
\definecolor{mPurple}{rgb}{0.58,0,0.82}
\definecolor{backgroundColour}{rgb}{255,255,255}
\definecolor{blue}{rgb}{0, 0, 1}

\lstdefinestyle{CStyle}{
    backgroundcolor=\color{backgroundColour},   
    commentstyle=\color{mGreen},
    keywordstyle=\color{magenta},
    numberstyle=\tiny\color{mGray},
    stringstyle=\color{mPurple},
    basicstyle=\footnotesize,
    breakatwhitespace=false,         
    breaklines=true,                 
    captionpos=b,                    
    keepspaces=true,                 
    numbers=left,                    
    numbersep=5pt,                  
    showspaces=false,                
    showstringspaces=false,
    showtabs=false,                  
    tabsize=2,
    language=C
}

\lstdefinestyle{PyStyle}{
    backgroundcolor=\color{backgroundColour},   
    commentstyle=\color{mGreen},
    keywordstyle=\color{magenta},
    numberstyle=\tiny\color{mGray},
    stringstyle=\color{mPurple},
    basicstyle=\footnotesize,
    breakatwhitespace=false,         
    breaklines=true,                 
    captionpos=b,                    
    keepspaces=true,                 
    numbers=left,                    
    numbersep=5pt,                  
    showspaces=false,                
    showstringspaces=false,
    showtabs=false,                  
    tabsize=2,
    language=Python
}
\journal{Computer Physics Communications}

\begin{document}

\begin{frontmatter}



\title{pyLOM: A HPC open source reduced order model suite for fluid dynamics applications}

\affiliation[inst1]{
            organization={Barcelona Supercomputing Center},
            addressline={Plaça Eusebi Güell, 1-3}, 
            city={Barcelona},
            postcode={08034}, 
            state={Barcelona},
            country={Spain}            
}

\affiliation[inst2]{
            organization={Universitat Politècnica de Catalunya},
            addressline={Carrer Colom, 11}, 
            city={Terrassa},
            postcode={08222}, 
            state={Barcelona},
            country={Spain} 
}

\affiliation[inst3]{
            organization={Universidad Politécnica de Madrid},
            addressline={Plaza del Cardenal Cisneros, 3}, 
            city={Madrid},
            postcode={28040}, 
            state={Madrid},
            country={Spain} 
}

\author[inst1,inst2]{Benet Eiximeno}
\author[inst1]{Arnau Miró}
\author[inst3]{Beka Begiashvili}
\author[inst3]{Eusebio Valero}
\author[inst2]{Ivette Rodriguez}
\author[inst1]{Oriol Lehmkhul}

\begin{abstract}
 This paper describes the numerical implementation in a high-performance computing environment of an open-source library for model order reduction in fluid dynamics. This library, called pyLOM, contains the algorithms of proper orthogonal decomposition (POD), dynamic mode decomposition (DMD) and spectral proper orthogonal decomposition (SPOD), as well as, efficient SVD and matrix-matrix multiplication, all of them tailored for supercomputers. 
 % Aqui pot portar bé parlar de l'escalabilitat
The library is profiled in detail under the MareNostrum IV supercomputer. The bottleneck is found to be in the QR factorization, which has been solved by an efficient binary tree communications pattern. Strong and weak scalability benchmarks reveal that the serial part (i.e., the part of the code that cannot be parallelized) of these algorithms is under 10\% for the strong scaling and under 0.7\% for the weak scaling.
% Ara li donem punch amb el exemple
Using pyLOM, a POD of a dataset containing $1.14\times 10^8$ gridpoints and 1808 snapshots that takes 6.3Tb of memory can be computed in 81.08 seconds using 10368 CPUs.
Additioally, the algorithms are validated using the datasets of a flow around a circular cylinder at $Re_D = 100$ and $Re_D = 1\times 10^4$, as well as the flow in the Stanford diffuser at $Re_h = 1\times 10^4$. 
\end{abstract}




\end{frontmatter}


\section{Introduction}
\label{sec:intro}
% Benet
Modal decomposition techniques are widely used to reduce the complexity of complex mathematical models. They are of interest in all the fields that involve the treatment of large amounts of data, such as in medical engineering \citep{grimberg2000examination}, signal analysis \citep{hinze2012discrete}, structural engineering \citep{kappagantu1999optimal, kappagantu2001analysis} or fluid dynamics \citep{berkooz_proper_nodate}.

These decompositions identify the dominant features in the system and give information about their relevance and dynamics. Moreover, they allow the reconstruction of the database using only the most meaningful features for the study, either related to a particular frequency or to a certain level of energy contained in the structures. This reconstruction enables the filtering of the noise and small fluctuations of the information.

A clear advantage of working with only the most important features of the system is the possibility of identifying which of them is responsible for every pattern in the behavior of the model and removing the modes that are detrimental to the system. Furthermore, their capacity to filter out the noise makes modal decompositions a good tool for predicting snapshots that are not included in the original database.

The most classical decomposition used in the field of fluid dynamics is proper orthogonal decomposition (POD). POD was first introduced in fluid dynamics by Lumley \cite{lumley_rational_2004} with the attempt to decompose the randomness of turbulence into modes that have some portion of the total fluctuating kinetic energy of the flow. Sirovich \cite{sirovich} explored the relationship between that decomposition and the coherent structures of the flow, making POD a relevant tool for the study of vortex dynamics in all types of fluid flows. For instance, Del Pino et al. \cite{del2011dynamics} studied the dynamics of the wing tip vortex of a NACA 0012 airfoil with the POD techniques developed by Roy and Leweke \cite{roy2008experiments}, and Zhang et al. analyzed the flow characteristics in a centrifugal pump \cite{zhang2021unsteady}.

Besides giving new insight into the data analysis, POD in fluid dynamics is also helpful to create a surrogate model of the case by projecting the equations to the most energetic modes \citep{smith2005low, willcox2002balanced, yondo_review_2018, wei2022parametric} or by learning the evolution of the modes with artificial neural networks \cite{colanera2data, calzolari2022deep, oulghelou2022surrogate}.

Recently, dynamic mode decomposition (DMD) and spectral proper orthogonal decomposition (SPOD) have been introduced as an improvement of POD. The biggest difference between them is that each POD mode is associated with a temporal signal that can have several dominant frequencies, while DMD and SPOD modes are related to a single frequency. Moreover, POD and SPOD order the modes regarding their physical energy, while DMD classifies the modes concerning their dynamical importance to minimize errors in the reconstruction of the flow field.

DMD was introduced by Schmid \cite{schmid2010dynamic} to have a frequency-based modal decomposition for fluid dynamics. In DMD, the dominant frequencies are detected and associated with the spatial structures through the eigenvalues and eigenvectors of an approximate inter-snapshot linear mapping \citep{schmid2011application}.

DMD has been continuously updated in recent years as some corrections have been presented for cases like transonic flows \citep{jiaqing2018dynamic}, active flow control applications \citep{proctor2016dynamic} or general datasets without constant timestep between snapshots \citep{dawson2016characterizing}. A first summary of the initial theoretical basis and applications can be found in Kutz et al. \cite{kutz2016dynamic}, while a more updated review of the DMD methodologies was done by Schmid in 2022 \cite{schmid2022dynamic}.

The investigation by Garicano-Mena et al. \cite{garicano2019composite} and Li et al. \cite{li2020dynamic} of an actuated turbulent channel to explore the existence of flow features linked to drag reduction, is a good example of the DMD usage for coherent structures characterization. Another example is the study from Barros et al. \cite{barros2022dynamic}, who used DMD to extract features from observations with different mesh topologies and dimensions as the ones found in adaptive mesh refinement simulations. Moreover, DMD is also useful to obtain snapshots that have not been computed or that are outside the simulated interval without using any additional technique to create a surrogate model \citep{yuan2021flow}.

SPOD was presented by Towne et al. \cite{towne2018spectral} as an improvement of POD when the relevant motions occur at low energies or multiple frequencies. Similarly to POD and DMD, SPOD can be used for both identification of coherent structures and low-rank reconstruction. For instance, Karami and Soria \cite{karami2018analysis} employed SPOD to study the spatiotemporal dynamics of an under-expanded supersonic impinging jet. Abreu et al. \cite{abreu2020spectral} also used SPOD to identify energetically dominant coherent structures in turbulent pipe flows. On the other hand, Nekkanti and Schmidt \cite{nekkanti2021frequency} used large-eddy simulation data of a turbulent jet to demonstrate the applicability of the SPOD algorithm for low-rank reconstruction and denoising.

The popularity and broadband usage of these three modal decompositions in fluid dynamics led to the publication of several reviews that compare the capabilities of the most relevant modal decompositions (see for instance Begiashvili et al. \cite{begiashvili2023data}) and various open-source libraries to compute POD, DMD and SPOD. For example, pykoopman \cite{pan2023pykoopman} and PyDMD \cite{demo2018pydmd} have serial and non-compiled implementations of several DMD variants for model order reduction. Moreover, there are some Python-based parallel implementations for modal decompositions as PyParSVD \cite{maulik2021pyparsvd}, which contains the distributed, streaming and randomized versions of the single-value decomposition for POD and PySPOD \cite{mengaldo2021pyspod} which has a parallel implementation of the SPOD.

The library developed in this work is named pyLOM, which stands for Python low-order modeling. pyLOM is completely open source and can be downloaded from its github repository \cite{pyLOM}. Up to the authors' knowledge, pyLOM is the first library to include the three algorithms in a high-performance computing environment. The aim of this paper is to detail the parallel implementation of these algorithms as well as the analysis of the scalability and efficiency of each of their components.

The remainder of the manuscript is organized as follows: \autoref{sec:math} explains the mathematical nuances of each algorithm, \autoref{sec:num} details the implementation and parallelization of the decompositions in pyLOM, \autoref{sec:validation} proves the accuracy of the implementations and finally \autoref{sec:performance} provides a detailed analysis of the strong and weak scalability of each decomposition.

\section{Mathematical description}
\label{sec:math}
\subsection{Proper orthogonal decomposition}
Proper orthogonal decomposition captures an infinite-dimensional process with a reduced number of modes \citep{holmes_low-dimensional_1997}. This method is based on finding the vectors of a basis to decompose a field $F(X, t)$ into a set of deterministic functions that characterize the dominant features of the system. This decomposition can be written as
\begin{equation}
    F(X, t) = \sum_{i = 1}^{i = N}a_i(t) \Phi_i(X),
    \label{eqn:POD}
\end{equation}
where $N$ is the number of functions to decompose the field in.
%Each of the modes is linked to a spatial correlation and a temporal coefficient.
Since the definitions of the time coefficients, $a_i(t)$, and the spatial modes, $\Phi(X)$, are not unique, the basis for the spatial modes is required to be orthonormal, i.e., 
\begin{equation}
\int_{\mathbf{X}} \Phi_{i_1}(X) \Phi_{i_2}(X) \mathrm{d} x=\left\{\begin{array}{l}
1 \text { if } i_{1}=i_{2} \\
0 \text { otherwise }
\end{array}\right.
\end{equation}

In addition to orthonormality, the chosen vectors to build the basis must be optimal. To do so, the vectors are ordered so that the first $N_r$ vectors are the ones that reconstruct the database with the minimum possible error using $N_r$ modes. 

Spatial POD modes, $\Phi(X)$, can also be seen as the set of deterministic functions that best approximate a stochastic function $F(X,t)$ on average. Under this definition, the computation of the vectors in the orthonormal basis is formally defined as the maximization of
\begin{equation}
    \centering
    \label{eqn:fumada}
    \lambda = \frac{E\left\{\left| \left< F(X,t), \Phi(X) \right> \right|^2\right\}}{\left< \Phi(X), \Phi(X) \right>}
\end{equation}
where $E\left\{\cdot \right\}$ is the expectation operator over the probability space. The function $\Phi(X)$ must then satisfy the eigenvalue problem 
\begin{equation}
    \label{eqn:fumpod}
    \centering
    \left<C(X, X'), \Phi(X')\right> = \lambda \Phi(X)
\end{equation}
where $C(X, X')$ is the two-point spatial correlation tensor. The orthogonality and optimality properties that characterize the POD come from this tensor being a nuclear kernel.
    
\subsection{Dynamic mode decomposition}
The first formulations of the dynamic mode decomposition were presented by Schmid \cite{schmid2010dynamic} and Rowley et al. \cite{ROWLEY_MEZIĆ_BAGHERI_SCHLATTER_HENNINGSON_2009}. It arises from the need to have a snapshot-based decomposition which results in modes that are directly related to the coherent structures of the flow. 

The results of DMD can be interpreted as structures of a linear tangent approximation to the underlying flow and describe fluid elements that have a dominant dynamic behavior inside the captured data. The fact that the modes extracted from DMD are part of a dynamic system implies that they are not correlated with a time signal anymore but with a coherent structure and an associated frequency, amplitude and damping ratio. 

The data used for DMD has to be constantly sampled in time ($\Delta t = ct.$) and has to be presented in a snapshot sequence given by the matrix ${\cal{D}}$,
\begin{equation}
    {\cal{D}} = \left[d_1,d_2,d_3,...,d_N\right]
\end{equation}
where $d_i$ stands for the ith sample of the flow field. The first step is to assume that it exists a linear mapping $A$ which connects the field $d_i$ to the next field $d_{i+1}$,
\begin{equation}
    d_{i+1} = A d_i\text{,}
\end{equation}
if $A$ is assumed as constant (Koopman assumption), it is possible to  formulate the sequence of flow fields as a Krylov sequence:
\begin{equation}
    \label{eqn:krylov}
    {\cal{D}} = \left[d_1, Ad_1, A^2d_1,...,A^{N-1}d_1\right]
\end{equation}

The goal of DMD is to find the linear operator $A$ and then compute the dynamic characteristics of the dynamical process described by $A$ based on the sequence ${\cal{D}}$. In this work, the chosen approach for the computation of $A$ is the algorithm presented by Tu et al. \cite{tu2014clarence}, further discussed in \autoref{sec:num}.

The dynamic information is computed with the eigendecomposition of $A$ and the frequency of the modes is obtained as
\begin{equation}
\label{eqn:freq}
    f_i = \frac{\theta_i}{2\Delta t \pi},
\end{equation}
where $\theta_i$ is the argument of the eigenvalue $\mu_i$. 

The relevance that DMD modes have in the flow is given by their amplitude. Among the different methods presented to rank the DMD modes, the most efficient and the one implemented in pyLOM, is the method introduced by Jovanovic et al. \cite{jovanovic2014sparsity}.

\subsection{Spectral proper orthogonal decomposition}
Spectral proper orthogonal decomposition is an extension of POD presented by Towne et al. \cite{towne2018spectral} that aims to seek spatio-temporal correlations. It bases its analysis on a stochastic ensemble that consists of a series of repetitions of the time-dependent flow. 

SPOD modes are the ones that best approximate $F(X,t)$ on average, therefore, the quantity to maximize is now
\begin{equation}
    \centering
    \lambda = \frac{E\left\{\left| \left< F(X,t), \Phi(X,t) \right> \right|^2\right\}}{\left< \Phi(X,t), \Phi(X,t) \right>},
\end{equation}
which leads to the eigenvalue problem 
\begin{equation}
    \centering
    \left<C(X, X', t, t'), \Phi(X',t')\right> = \lambda \Phi(X,t),
\end{equation}
where $C(X, X', t, t')$ is the two-point space-time correlation tensor. In contrast to the two-point space correlation tensor of POD, this tensor is not a nuclear kernel because statistically stationary flows persist indefinitely and have infinite energy in a space-time norm.  

A solution to this is working in the frequency domain and finding modes with the spectral eigenvalue problem. According to Towne et al. \cite{towne2018spectral}, the cross-spectral density tensor is defined as the Fourier transform of the difference between two timesteps of the correlation tensor
\begin{equation}
    \label{eqn:csd}
    \boldsymbol{S}\left(X, X^{\prime}, f\right)=\int_{-\infty}^{\infty} \boldsymbol{C}\left(X, X^{\prime}, \tau\right) \mathrm{e}^{-\mathrm{i} 2 \pi f \tau} \mathrm{d} \tau,
\end{equation}
where $\tau = t - t'$. The cross-spectral density tensor is nuclear, so at each frequency, there is a finite set of modes, $\psi_j(X, f)$, that are orthogonal to all other modes at the same frequency. Hence, for stationary flows, spectral POD modes oscillate at a single frequency and optimally represent the second-order space-time flow statistics.

\section{Numerical implementation}
\label{sec:num}
% Benet
This section describes how pyLOM utilizes high-performance computing techniques to efficiently compute POD, DMD, and SPOD of very large datasets. The front end of pyLOM is a Python package that calls a set of built-in functions to compute the modal decompositions and manage the input-output of the data in parallel. All computations are written using in C and then  wrapped in Cython \cite{cython}. Moreover, all the matricial algebraic operations are done with optimized libraries such as BLAS \cite{dongarra1990set}, LAPACK \cite{anderson1999lapack} or Intel's MKL (Math Kernel Library).

\begin{figure}
  \centering
  \includegraphics[width=\textwidth]{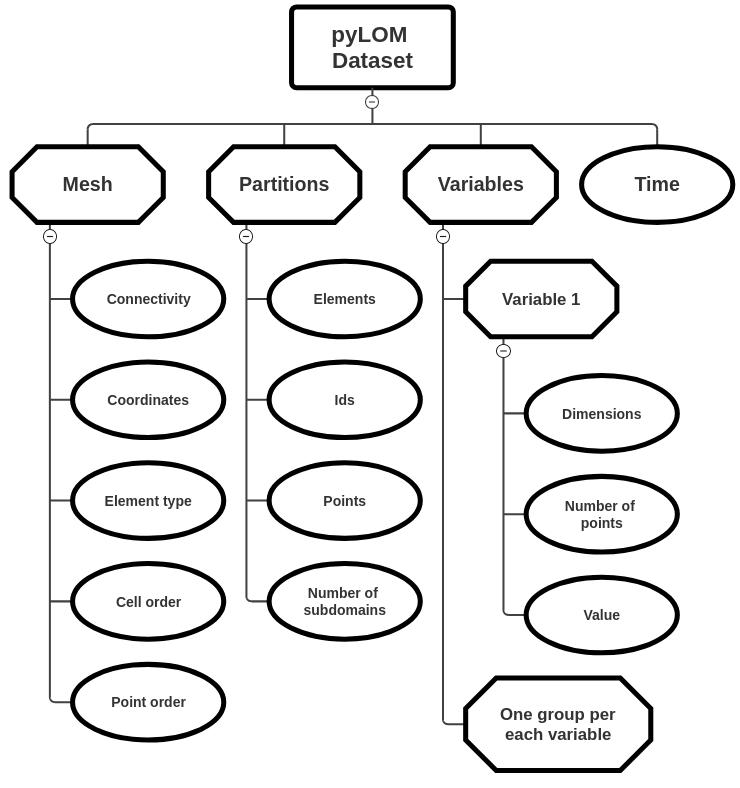}
  \caption{Hierarchical structure of the HDF5 pyLOM dataset file. Groups are identified by an octagon and datasets with an elipse}
  \label{fig:dataset}
\end{figure}

The input data is organized in snapshot matrices, $\cal{D}$, of size $M \times N$, where $M$ is the number of points and $N$ is the number of snapshots. These arrays and the rest of information needed to compute and visualize the modal decompositions are saved in HDF5 files that follow the hierarchical structure described in \autoref{fig:dataset}. This format allows 
%to have a hierarchical structure of the data, 
cross-platform compatibility, support for various data types, parallel I/O support and efficient chunking and compression capabilities.

Typically in reduced order model applications, the number of points of the computational mesh $M$ is much larger than the number of snapshots saved $N$. Hence, $\cal{D}$ is assumed as a tall and skinny matrix (i.e., $M >>> N$). To ensure that the array fits in the available memory, the data is distributed across the available processors by splitting $\cal{D}$ in arrays of size $M_i \times N$, where $M_i$ is the number of rows read by each processor. Henceforth, ${\cal{D}}_i$ stands for the local data array in each processor. The parallel implementation in pyLOM is done through message passing interface (MPI) \citep{Walker1996MPI:Interface}, which allows to use both distributed and shared memory machines.

\subsection{Proper orthogonal decomposition implementation}
Among the different methods to compute the eigendecomposition in \autoref{eqn:fumpod} (see for instance the principal component analysis \cite{hotelling1933analysis} and the Karhunen-Loève decomposition \cite{murdoch_linear_1970}), the chosen algorithm in pyLOM is the single value decomposition (SVD), which is an extension of the eigenvalue decomposition for non-square matrices. 

The SVD decomposes the initial snapshot matrix into the left singular vectors, $U$, the singular values, $S$, and the right singular vectors, $V$,
\begin{equation}
    {\cal{D}} = USV^{T}.
    \label{eqn:SVD}
\end{equation}
 
The size of $U$ is the same as the size of $\cal{D}$, $M\times N$, and each column contains a singular vector for all points of the domain. When looking back to \autoref{eqn:POD}, $U$ is directly linked with the spatial modes, $\Phi(X)$, as it only depends on space. 

The singular values are given in a $N\times N$ diagonal matrix containing the energy contribution of each singular vector in descendent order. They can be treated as the energy of the modes. The higher the singular value, the more energy is contained in the mode. The accumulated energy of the mode, relative to the total energy of the system is computed as:
\begin{equation}
    \frac{\sum_{i = 1}^{i=mod}S^2_i}{\sum_{i = 1}^{i=N}S^2_i}
\end{equation}

The right singular vectors, $V$, are a matrix of size $N\times N$ and they only depend on time. As the SVD gives the transposed right singular vectors, each row of $V$ is the time coefficient, $a_i(t)$, of one mode. 

Chan et al. \cite{chan1982improved} identified that for tall and skinny matrices it is more efficient to compute the SVD by expressing  ${\cal{D}}$ as the product between an orthogonal matrix, $Q$ ($M\times N$), and an upper triangular matrix, $R$ ($N\times N$) with the QR factorization \citep{trefethen1997numerical}, 
\begin{equation}
  \label{eqn:svdqr1}
  {\cal{D}} = Q\left(\begin{array}{l}R\\0\end{array}\right),
 \end{equation}
and then apply the single value decomposition only to the upper triangular matrix R,
 \begin{equation}
  \label{eqn:svdqr2}
  \left(\begin{array}{l}R\\0\end{array}\right) = U_rSV^T
 \end{equation}

 Finally, the left singular vectors are computed as 
  \begin{equation}
  \label{eqn:svdqr3}
  U= QU_r
 \end{equation}

In pyLOM, the QR decomposition is performed following the parallel algorithm presented in Demmel et al. \cite{demmel_communication-optimal_2012} for tall and skinny matrices. First of all each processor factorizes their local data array, ${\cal{D}}_i$, in $Q_{1i}$ and $R_i$. Then, all the values of $R_i$ are reduced to a single rank that computes the global value of $R$ and broadcasts it back so that each processor can compute its chunk of $Q$. 

The reduction and broadcasting operations are based on a binary tree algorithm. This communication scheme avoids a simultaneous communication of all cores with a single rank. Instead, there are $n = \log_2{\left(2^{\lceil \log_2 (P) \rceil}\right)}$ communication levels, where $P$ is the number of processors. All processors involved in each communication level, $i_{level}$, store their value of $R$ in the upper part of a buffer array $C$ of size $(2N;N)$. Each rank finds the processor to communicate with through a bitwise \texttt{XOR} between the processor number and the communication level, $i_{rank}\text{ \textasciicircum } i_{level}$. A bitwise \texttt{AND} between the processor and the communication level, $i_{rank}\text{ \& } i_{level}$, is used to alternate the role of sender (true) and receiver (false). In the latter case, the received $R$ is stored in the lower part of the buffer array $C$. Before entering the next reduction level, the receivers compute the $QR$ decomposition of $C$ to obtain $Q_{2i}$ and update $R$.

A similar methodology is used for the broadcasting of $R$ and $Q_{2i}$ so that each processor has its chunk of the orthogonal Q array, $Q_i$, and they all share the upper triangular matrix $R$. The information is sent through the buffer array $C$ which has $R$ in its upper part and $Q_{2i}$ in its lower. At each communication level $Q_{2i}$ is updated with the product between $C$ and $Q_W$, the $Q_{2i}$ received in the last communication level (if it the processor has not received any data during the broadcasting process, $Q_W$ is the identity matrix). At the end of the broadcasting process $Q_i$ is computed as the product between $Q_{1i}$ and the last $Q_W$ in each rank.

According to Demmel et al. \cite{demmel_communication-optimal_2012}, the total run-time $t$ of the algorithm is
\begin{equation}
    \centering
    t = \alpha \times \text{FLOPs} + \beta \times \text{data} + \gamma \times \text{messages},
    \label{eqn:time}
\end{equation}
where $\alpha$ is the time per FLOP, $\beta$ is the inverse of the bandwidth and $\gamma$ is the latency. The number of FLOPs for this algorithm is
\begin{equation}
    \centering
    \text{FLOPs} = \frac{2MN^2}{P} + \frac{2N^3}{3} \log{P},
    \label{eqn:flops}
\end{equation}
where first term is the number of FLOPs needed to QR factorize the initial matrix $\cal{D}$ and the second term is the number of operations involved in the subsequent QR factorizations of the algorithm. The volume of data exchanged between cores is
\begin{equation}
    \centering
    \text{data} = \frac{N^2}{2}\log{P},
    \label{eqn:data}
\end{equation}
and the number of communicated messages is
\begin{equation}
    \centering
    \text{messages} = \log{P}.
    \label{eqn:messages}
\end{equation}

This information can be included in \autoref{eqn:time}, 
%\begin{equation}
%    \centering
%    t = \left(\frac{2MN^2}{P} + \frac{2N^3}{3}\log(P)\right)\alpha + \frac{N^2}{2}\log(P)\beta + \gamma \log(P)
%\end{equation}
and then rearranged to separate the effect of the QR decomposition of the initial matrix $\cal{D}$ and the time used in the communication process
\begin{equation}
    \centering
    t = \frac{2MN^2}{P}\alpha + \left(\frac{2N^3}{3}\alpha + \frac{N^2}{2}\beta + \gamma\right) \log(P).
    \label{eqn:time2}
\end{equation}

To complete the SVD, each processor has to compute locally the single value decomposition of  $R$ (\autoref{eqn:svdqr2}) and adjust the left singular vectors as in \autoref{eqn:svdqr3}. With this strategy, the left singular vectors, $U_i$, have the same partition as the input matrix, ${\cal{D}}_i$, whereas the singular values, $S$, and the right singular vectors, $V$, are common in all processors.

The reader is referred to \ref{sec:qr} for a detailed algorithm of the implementation of the parallel QR decomposition (\autoref{alg:tsqr_svd}), as well as an example of the truth tables of the operations needed for reducing a case using 6 processors. The detailed algorithm for the proper orthogonal decomposition can be found in \ref{sec:pod}.

\subsection{Dynamic mode decomposition implementation}
DMD is about finding the dynamic response of a linear mapping, $A$, that allows the formulation of a sequence of flow fields as a Krylov sequence (\autoref{eqn:krylov}). One of the most efficient and accurate ways to compute $A$ is presented by Tu et al. \cite{tu2014clarence} and is known as the exact DMD. In this procedure, the data is first organized in
\begin{equation}
    Y_1 = \left[d_1, ..., d_{N-1}\right]
    \label{eqn:y1}
\end{equation}
\begin{equation}
    Y_2 = \left[d_2, ..., d_{N}\right]
    \label{eqn:y2}
\end{equation}
then the Krylov sequence can be expressed as
\begin{equation}
    Y_2 = A Y_1
\end{equation}
and the Koopman operator is computed as
\begin{equation}
    A = U^TY_2VS^{-1}
\end{equation}
where $U$, $V$ and $S$ are the result of a SVD applied to $Y_1$. As $Y_1$ is distributed across the processors, this step is done using the parallel algorithm presented in \autoref{alg:tsqr_svd}. 

The results of the SVD can be truncated to a certain number of modes, $N_r$, to reduce the size of the matrix for the following computations. The truncation residual should be the lowest possible as long as the resulting computations fit in the available memory. According to Li et al. \cite{li2022parametric}, the truncation of any mode, regardless of its energy level, can lead to the loss of relevant temporal dynamics that affect the accuracy of the DMD results. 

The computation of the linear mapping $A$ involves the parallel matrix-matrix multiplication $C=U^TY_2$ as the transposed left eigenvectors, $U^T$, and the data array, $Y_2$ are distributed across the processors. In this particular case, the shape of the matrices is $(N_r \times M) \times (M \times N)$. Considering the data distribution in pyLOM, the left array is split column-wise, while the right array is split row-wise. The resulting array of shape $N_r \times N$ can be shared by all processors because $N_r \leq N << M$. 

This hypothesis allows for a simpler approach compared to a standard parallel-parallel matrix product. The operation is simplified to the global sum of the local products between the chunk of matrices stored in each processor. The reduction and sum of the result is done using the \texttt{all\_reduce} function from the MPI library.

The computational time to perform the matrix-matrix product can also be modeled with \autoref{eqn:time}. In this case, the total number of operations is
\begin{equation}
    \centering
    \label{eqn:flopsmatmul}
    \text{FLOPs} = N_rN\left(\frac{2M}{P}-1\right) + N_rN(P-1),
\end{equation}
where the first term is the number of operations due to the matrix product in each processor, and the second one accounts for the sum of the local results obtained by every CPU. Regardless of the number of processors, $2M/P >> 1$, therefore \autoref{eqn:flopsmatmul} can be simplified to
\begin{equation}
    \centering
    \text{FLOPs} = \frac{2MN_rN}{P} + N_rN(P-1)
\end{equation}

The total amount of exchanged data per processor is
\begin{equation}
    \centering
    \text{data} = N_rN(P-1)
\end{equation}
and the number of messages
\begin{equation}
    \centering
    \text{messages} = P-1
\end{equation}
This information is included in \autoref{eqn:time} to obtain the following model for the computational time of the implemented matrix-matrix product
%\begin{equation}
%    \centering
%    t = \left(\frac{2MN_rN}{P}+N_rN(P-1)\right)\alpha + N_rN\left(P - 1\right)\beta +(P-1)\gamma
%    \label{eqn:timematmul}
%\end{equation}
%and then \autoref{eqn:timematmul} is 
and then rearranged into separate terms: one that decreases and one that increases linearly with the number of processors
\begin{equation}
    \centering
    t = \frac{2MN_rN\alpha}{P} + (N_rN\alpha+ N_rN\beta + \gamma) (P-1).
    \label{eqn:timematmul2}
\end{equation}

Afterwards, the linear mapping $A$ is shared between all the processors and has a size of $N_r \times N$. Hence, the rest of the computations can be done in serial by all the processors without concerns for performance or memory usage. The next step is the eigendecomposition of $A$, $A = \mu w$, and then the DMD modes are computed as:
\begin{equation}
    \label{eqn:dmdmodes}
    \varphi = Y_2 V^TS^{-1}w/\mu
\end{equation}

To compute the amplitudes as in Jovanovic et al. \cite{jovanovic2014sparsity}, the reconstruction of the snapshots is taken as
\begin{equation}
    Y_1 = \varphi D_{\alpha}V_{and}
\end{equation}
where $D_{\alpha}$ is a diagonal matrix containing the vector of amplitudes of each mode. $V_{and}$ is the Vandermonde matrix of size $N_r \times N$ and governs the time evolution of the dynamic modes of the first $N_r$ eigenvalues of $A$. Finally, the computation of $D_{\alpha}$ is done by minimizing the following Frobenius norm
\begin{equation}
\underset{\alpha}{\operatorname{minimize}}\left\|Y_1- \varphi D_{\alpha} V_{\text {and }}\right\|_{F}^{2}
\end{equation}

The detailed algorithms of the computation of the DMD modes, including the parallel matrix-matrix multiplication, and their amplitudes according to Jovanovic et al. \cite{jovanovic2014sparsity} can be found in \ref{sec:dmd}.

\subsection{Spectral proper orthogonal decomposition implementation}
The first step in the SPOD computation is estimating the cross-spectral density tensor of \autoref{eqn:csd} at a predefined set of frequencies, $f$. To do so, the data matrix ${\cal{D}}$ is subdivided into $nBlks$ blocks of snapshots. Every block contains $npwin$ snapshots and overlaps with its neighboring blocks by sharing the initial and final $nolap$ snapshots.

The Fourier transform is applied to all data points in each block. As suggested by Schmidt and Colonius \cite{schmidt2020guide}, all snapshots are multiplied by the Hamming window function to taper the ends of the sequence and reduce spectral leakage \cite{von1903handbook}. Now the columns of the blocks represent the result of the Fourier transform at a certain frequency, $i_{freq}$. The frequency axis in all blocks is the same if the snapshots are evenly sampled in time. Therefore, the columns of all blocks that represent $i_{freq}$ can be grouped in an array $Q_f$ of size $M_i \times nBlks$. The cross-spectral density at each frequency can be estimated as the average of the different components of $Q_f$ along the blocks,
\begin{equation}
    \label{eqn:csdimp}
    \centering
    S_f = \frac{1}{nBlks}Q_fQ_f^{H},
\end{equation}
where $Q_f^H$ is the complex conjugated and transposed of $Q_f$. Note that $Q_f$ is row-wise distributed ($M_i;N$), thus $Q_f^H$ becomes column-wise distributed ($N;M_i$). Therefore, the operation of \autoref{eqn:csdimp} implies a parallel product of two  matrices resulting in a matrix that is both row- and column-wise distributed $(M_i;M_i)$.

The implementation of this parallel matrix-matrix multiplication is far more complex than the one presented in \autoref{alg:matmul_parañ} because each processor needs to communicate twice with all the other processors and the resulting array has to be assembled and redistributed both in rows and columns to ensure that it fits in memory.

Instead of following the original implementation of the SPOD by Towne et al. \cite{towne2018spectral}, pyLOM extracts the spatial modes and their energy with the procedure described by Frame and Towne \cite{frame2023space}. Their process is based on the single value decomposition of $Q_f$
\begin{equation}
    \centering
    \frac{1}{\sqrt{nBlks}} Q_f = U_f S_f V^H_f.
\end{equation}
Then, the modes and the energies are computed as
\begin{equation}
    \centering
    \Psi_f = U_f \text{ and } \Lambda_f = S_f^2.
\end{equation}

This implementation has the advantage of reusing the parallel SVD of \autoref{alg:tsqr_svd} and avoids computing and operating with the cross-spectral density tensor. Note that, at each frequency, there are a total of $nBlks$ modes that come ordered according to their energy. It is still needed to order the frequencies according to their highest energy mode.

\section{Validation}
\label{sec:validation}
The POD and DMD algorithms of pyLOM have already been used in several studies. For instance, Eiximeno et al. \cite{eiximeno2022wake} used POD to analyze the coherent structures and wake dynamics of a two-degree-of-freedom vibrating cylinder with a low mass ratio at a Reynolds number of $Re=5300$ \cite{pastrana2018large}. On the other hand, Montalà et al. \cite{richi} used POD to identify the streaks in the boundary layer of the main element of a 30p30n airfoil in high-lift configuration at a Reynolds number of $Re = 7.5\times 10^5$. Eiximeno et al. \cite{eiximeno2023hybrid} were the first ones to use the DMD implementation of pyLOM to classify the noise sources radiated by the airflow around a cylinder at a Reynolds number of $Re = 10^4$ and a Mach number of $M = 0.5$.

\begin{table}[]
    \centering
    \caption{Summary of the validation cases including the Reynolds number, the variables of the dataset, the number of grid points ($M$), the number of snapshots ($N$) and the time difference between them ($\Delta t$)}
    \label{tab:test_cases}
    \begin{tabularx}{\textwidth}{Xcccccc}
        \toprule
        \textbf{Case} & \textbf{Re} & \textbf{Variables} & \textbf{$M$} & \textbf{$N$} & \textbf{$\Delta t$} \\
        \midrule
        2D circular cylinder & 100 & $(u_x, u_y)$ & $1.78\times 10^5$ & 151 & 0.2 \\
        Stanford diffuser    & $1\times 10^4$ & $u_x$ & $1.14\times 10^8$ & 1808 & 0.43 \\
        3D circular cylinder & $1\times 10^4$ & $p$ & $2.73\times 10^7$ & 370 & 0.166 \\
        \bottomrule
    \end{tabularx}
\end{table}

At the moment of writing this manuscript, no published works used the SPOD algorithm of pyLOM. Its implementation is validated here with the test cases described in \autoref{tab:test_cases}. The first case is the decomposition of the flow around a circular cylinder at $Re_D= DU_{\infty}/\nu = 100$, where $D$ is the cylinder diameter, $U_{\infty}$ is the freestream velocity and $\nu$ is the fluid kinematic viscosity. The SPOD results are compared with the ones published in Begiashvili et al. \cite{begiashvili2023data} using 
%the Matlab SPOD library provided by Schmidt \cite{spod_matlab} 
another  SPOD implementation \cite{spod_matlab}.

For a fair comparison, the SPOD hyperparameters and the database used for the validation are identical to those used by Begiasvhili et al. \cite{begiashvili2023data}. The number of snapshots per window is set to $n_{win} = 60$, with an overlap of 50\% and the database is taken from Brunton and Kutz \cite{BruntonKutz2019}. 

\begin{figure}
    \centering
    \includegraphics[width=0.8\textwidth]{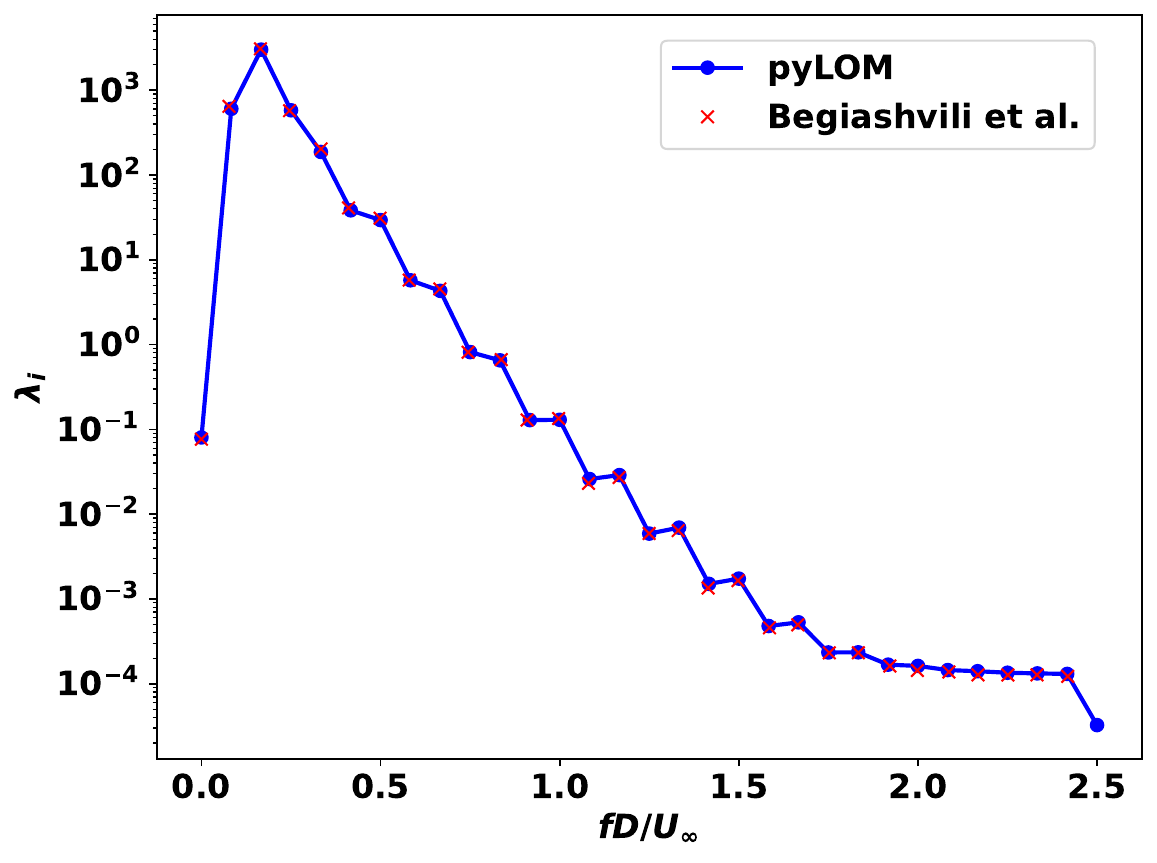}
    \caption{Comparison of the energy of each SPOD, $\lambda_i$, with the ones obtained by Begiashvili et al. \cite{begiashvili2023data} for the flow around a cylinder at $Re_D=DU_{\infty}/\nu=100$.}
    \label{fig:spec_spod}
\end{figure}

\begin{figure}
    \centering
    \subfloat[]{
        \includegraphics[width = 0.47\textwidth]{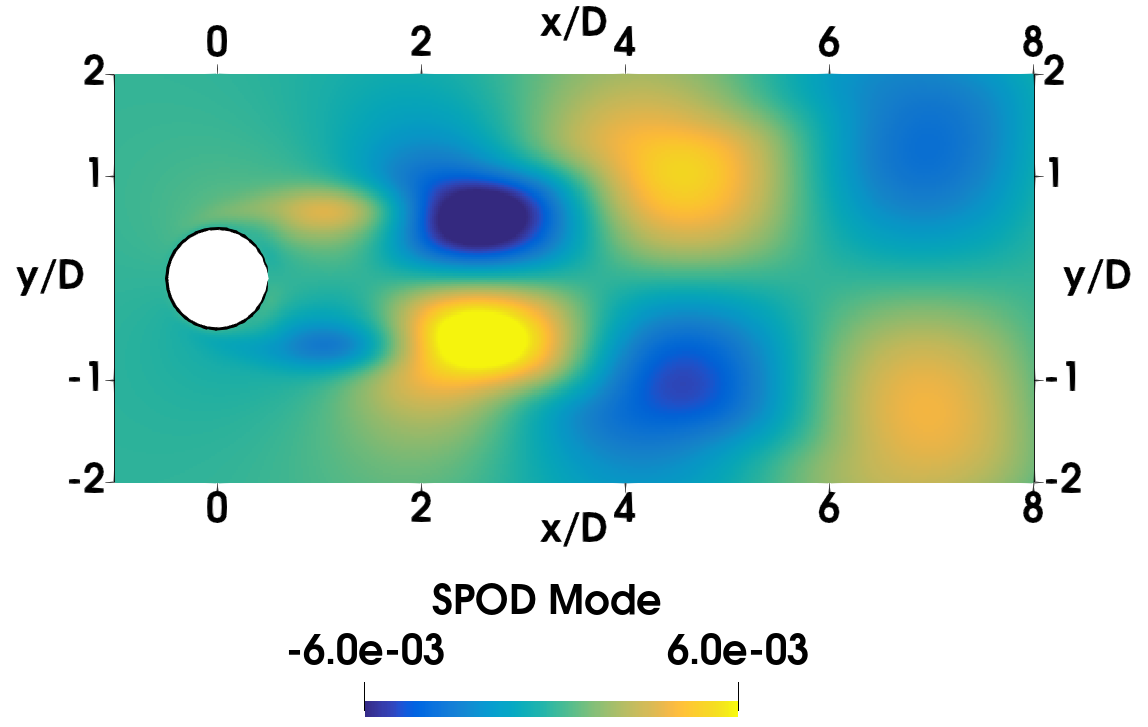}
        \label{fig:spod_cyl_1}
        }
        \subfloat[]{
        \includegraphics[width = 0.47\textwidth]{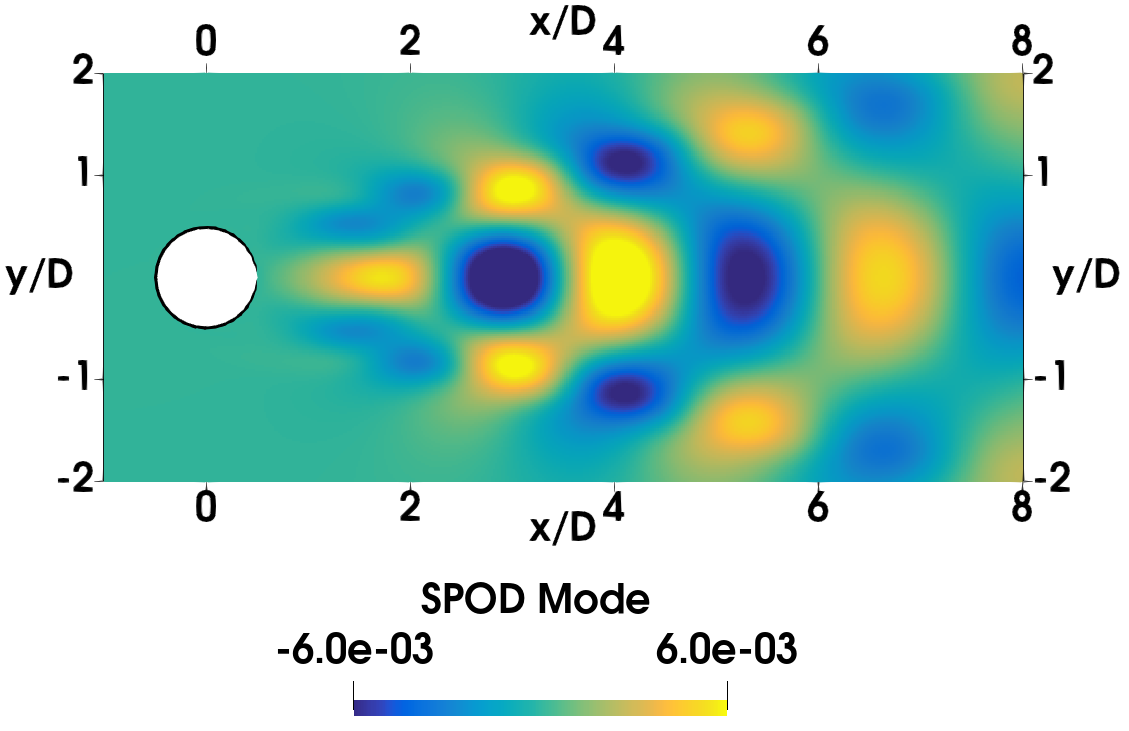}
        \label{fig:spod_cyl_2}
        }
        \caption{First (a) and second (b) most energetic SPOD modes of the airflow around a cylinder at $Re_D=DU_{\infty}/\nu=100$. They happen at the non-dimensional frequencies of $fD/U_{\infty}=0.166$ and $fD/U_{\infty}=0.333$, respectively}
        \label{fig:spod_cyl}
\end{figure}

\autoref{fig:spec_spod} shows that the energy spectrum of the SPOD analysis perfectly matches the results obtained by Begiashvili et al. \cite{begiashvili2023data}. Moreover, \autoref{fig:spod_cyl} shows the two most energetic spatial correlations, which happen at the non-dimensional frequencies of $fD/U_{\infty}=0.166$ and $fD/U_{\infty}=0.333$. They are similar to the ones presented in figure 37 from the paper published by Begiashvili et al. \cite{begiashvili2023data}, confirming that pyLOM has a correct implementation of the SPOD.

Then, the SPOD algorithm is tested in a high-performance computing environment by decomposing the flow in the Stanford diffuser at $Re_h=hU_b/\nu = 1 \times 10^4$, where $h$ is the height of the inlet channel of the diffuser, $U_b$ is the bulk inflow velocity and $\nu$ is the fluid kinematic viscosity. This database is the largest one studied so far with pyLOM, as it has $1.14\times 10^8$ points and 1808 snapshots, resulting in 6.3 Tb of memory. Miró et al. \cite{miro2023self} already computed the POD and DMD using pyLOM in 216 nodes of Marenostrum 4, in 81.08 and 947.89 seconds, respectively. The computations took 20.95 seconds for the SPOD.

\begin{figure}
\centering
\subfloat[]{\includegraphics[width=0.45\textwidth]{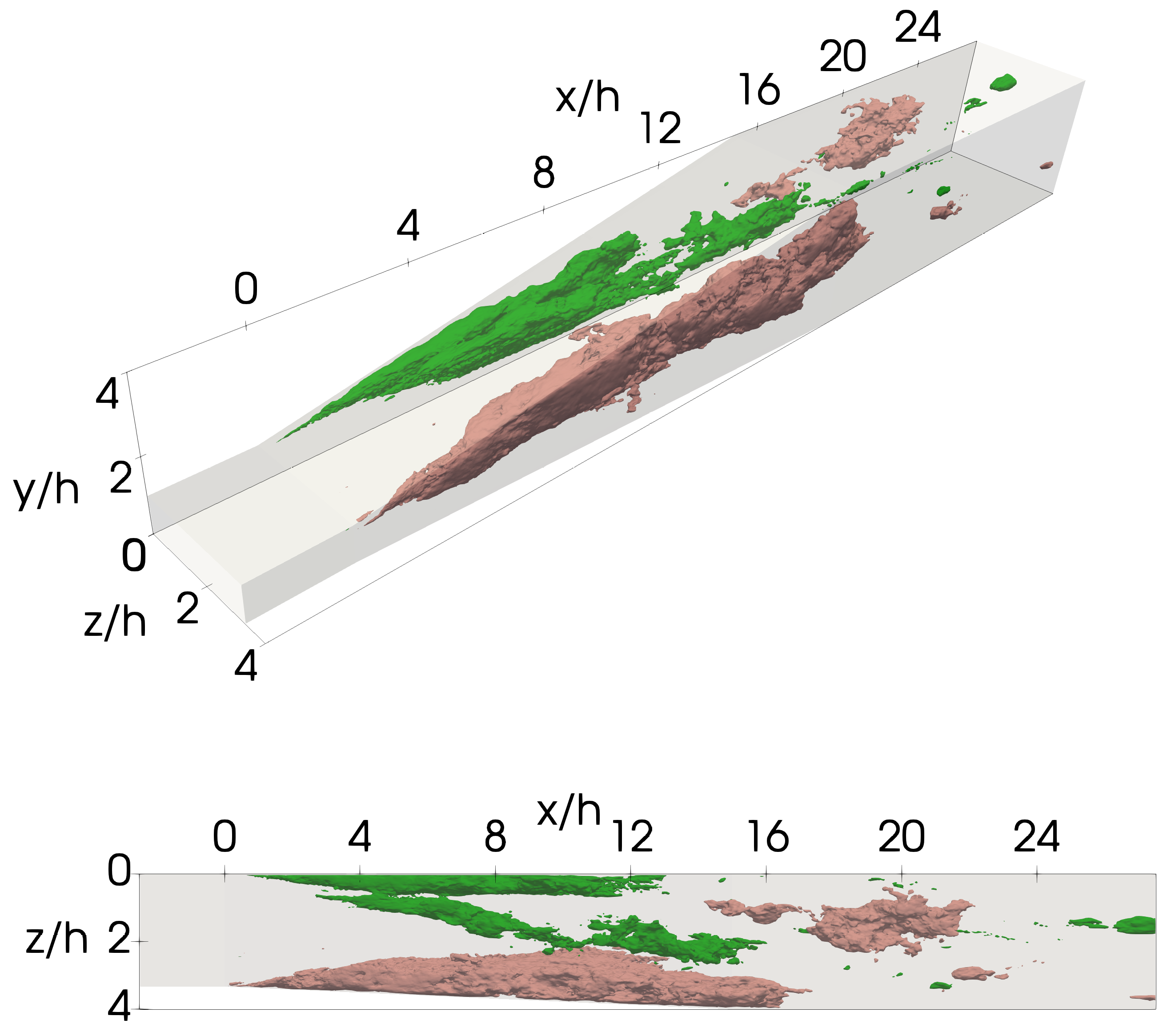} \label{fig:spoda}}
\subfloat[]{\includegraphics[width=0.45\textwidth]{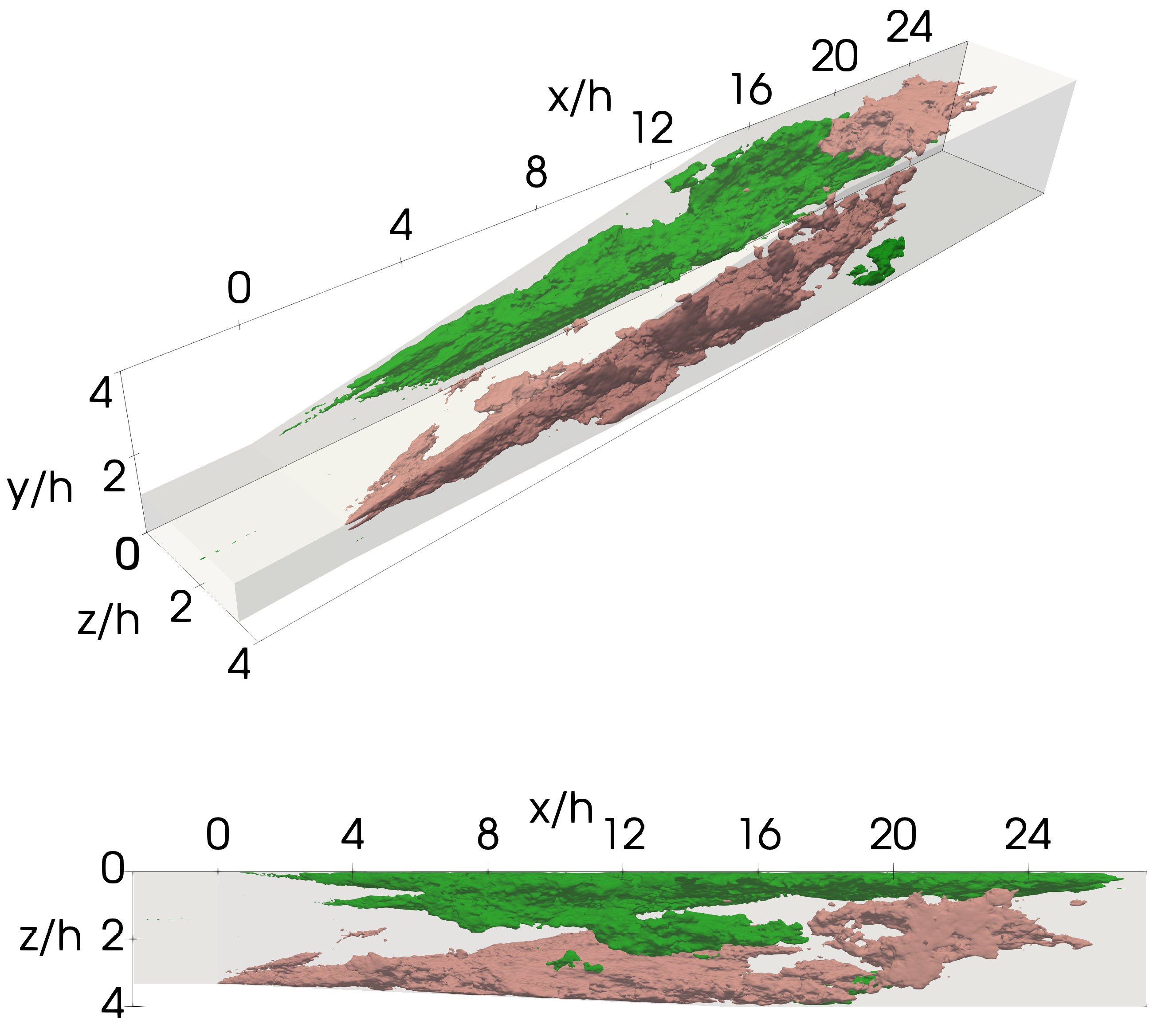}\label{fig:dmda}}\\
\subfloat[]{\includegraphics[width=0.45\textwidth]{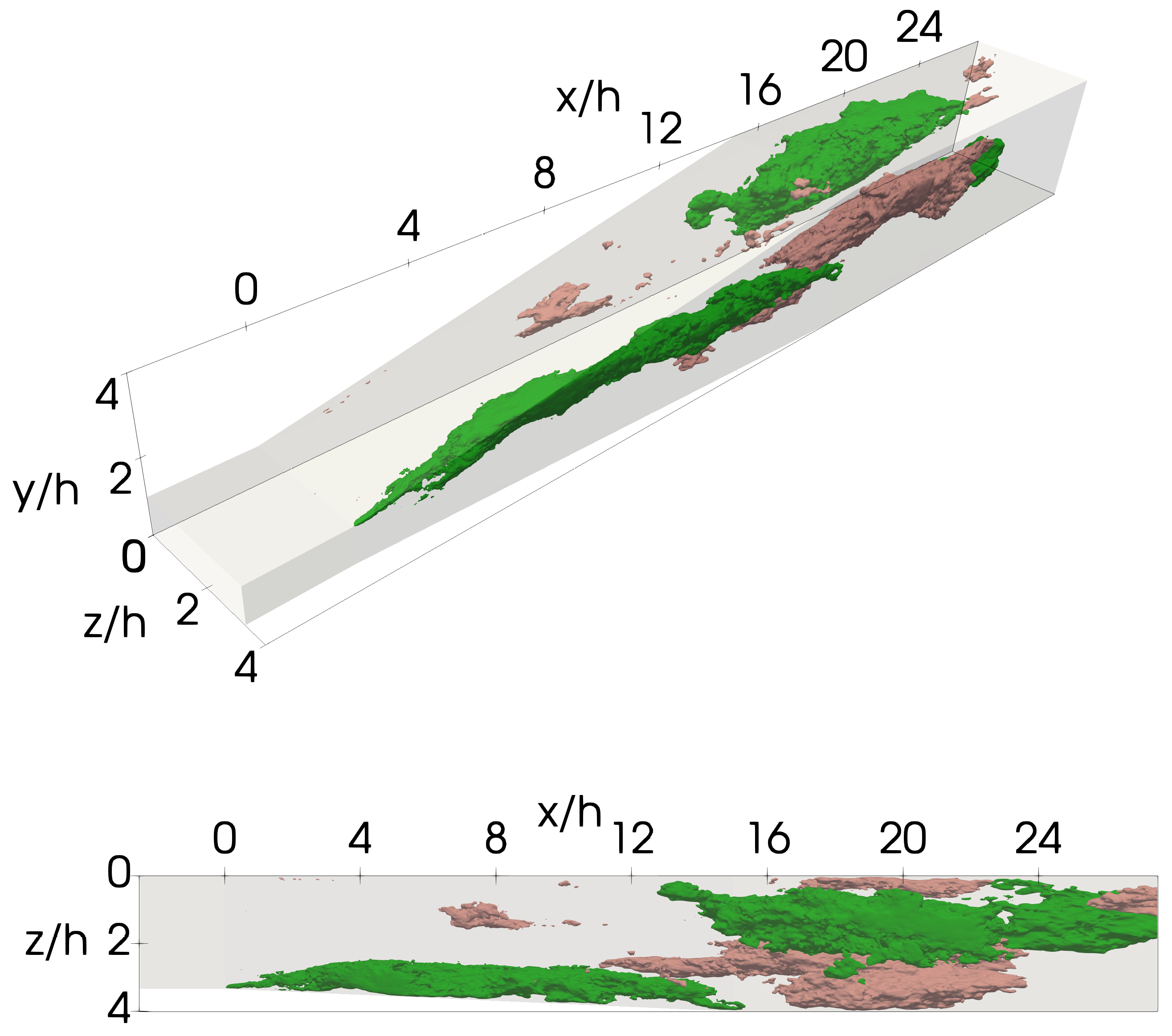}\label{fig:spodb}}
\subfloat[]{\includegraphics[width=0.45\textwidth]{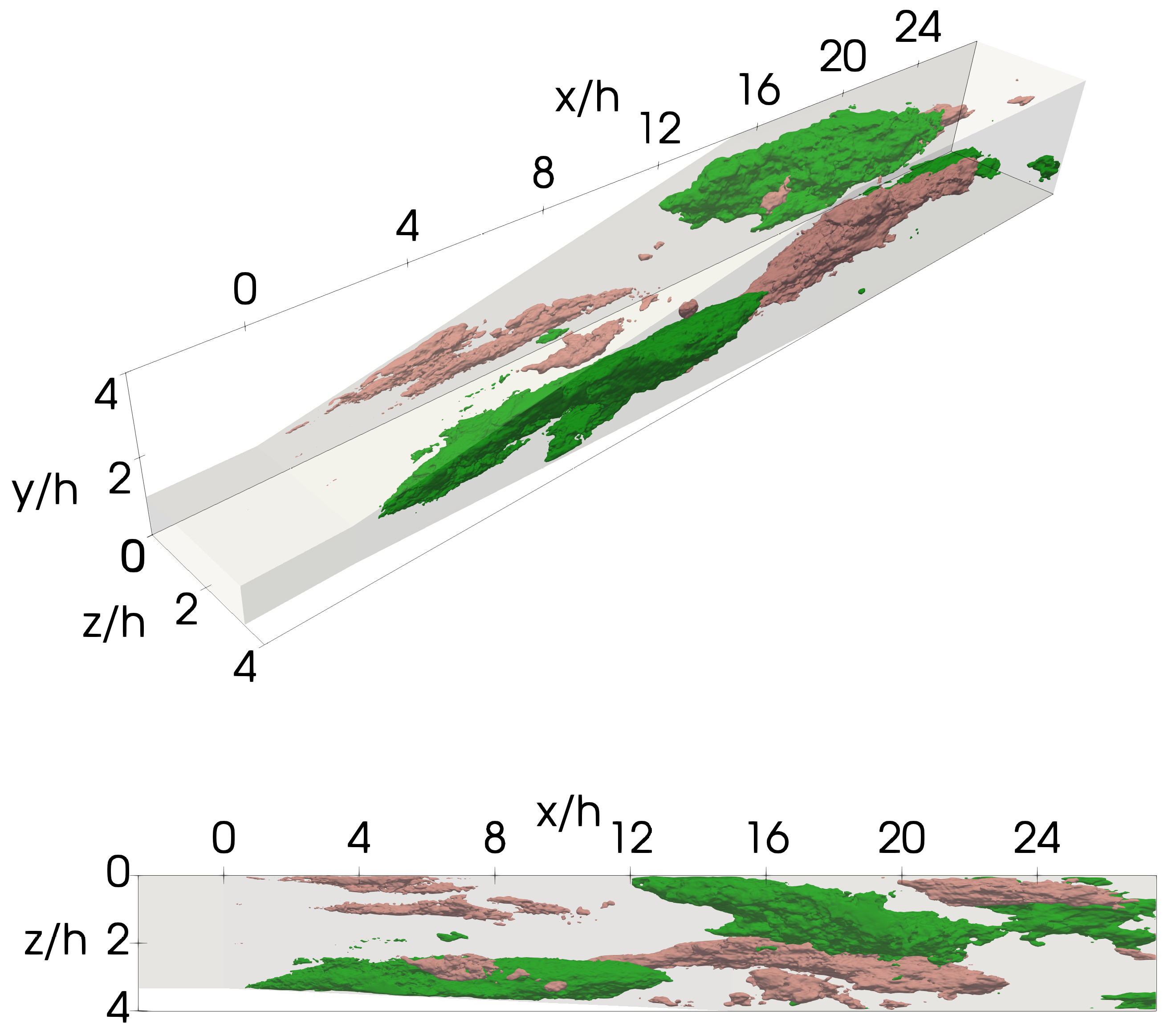}\label{fig:dmdb}}
\caption{Iso-contours for the SPOD (left) and DMD (right) modes of the spanwise velocity of the Stanford diffuser by Miró et al. \cite{miro2023self}. The associated frequencies to the structures depicted for the SPOD $fh/U_b=0.0038$ (a) and $fh/U_b=0.0076$ (c) and for the DMD $fh/U_b=0.0037$ (b) and $fh/U_b=0.0084$ (d)}\label{fig:Stan_SPOD_modes}
\end{figure}

The present SPOD results are compared with the DMD analysis presented by Miró et al.\cite{miro2023self}, which was used to study the large-scale motions originating in the top-right expansion corner of the Stanford diffuser. To do so, \autoref{fig:Stan_SPOD_modes} presents the two most relevant DMD and SPOD modes. The SPOD modes bear a striking resemblance with the already published data \cite{miro2023self}. In particular, the mode at the non-dimensional frequency of $fh/U_b=0.0038$ (\autoref{fig:spoda}) is very close to the DMD frequency of $fh/U_b=0.0037$ (\autoref{fig:dmda}) identified in Miró et al. \cite{miro2023self}. This structure corresponds to an overall acceleration-deceleration motion seen as a back-and-forth travelling wave that originates in the top-right expansion corner. Then, the mode at $fh/U_b=0.0076$ (\autoref{fig:spodb}) is also close to the frequency of $fh/U_b=0.0087$ (\autoref{fig:dmdb}) identified in Miró et al. \cite{miro2023self} and $fh/U_b=0.0084$ reported by Malm et al. \cite{malm2012coherent}. This frequency corresponds to a travelling wave, which produces a diagonal beating cross-stream motion \cite{miro2023self}.

%Apart from confirming a low-frequency diagonal cross-stream traveling wave \cite{malm2012coherent}, Miró et al. \cite{miro2023self} linked a second low-frequency to the continuous secondary movements, functioning as an accelerating and decelerating motion situated within the straight area of the diffuser. 

\begin{figure}
    \centering
    \subfloat[]{
        \includegraphics[width = 0.9\textwidth]{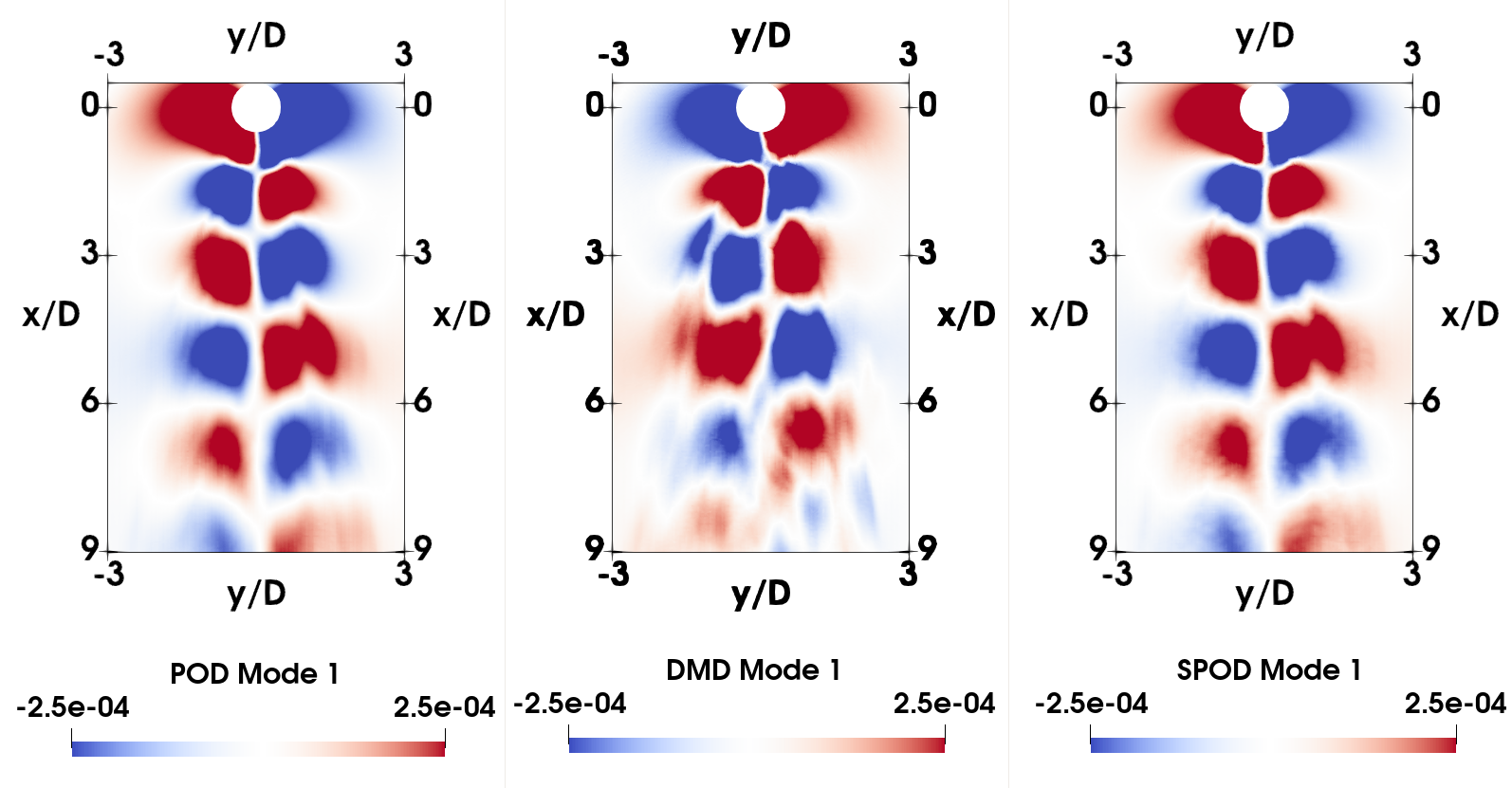}
        \label{fig:modecyl}
        }\\
        \vspace{0.3cm}
        \subfloat[]{
        \includegraphics[width = 0.9 \textwidth]{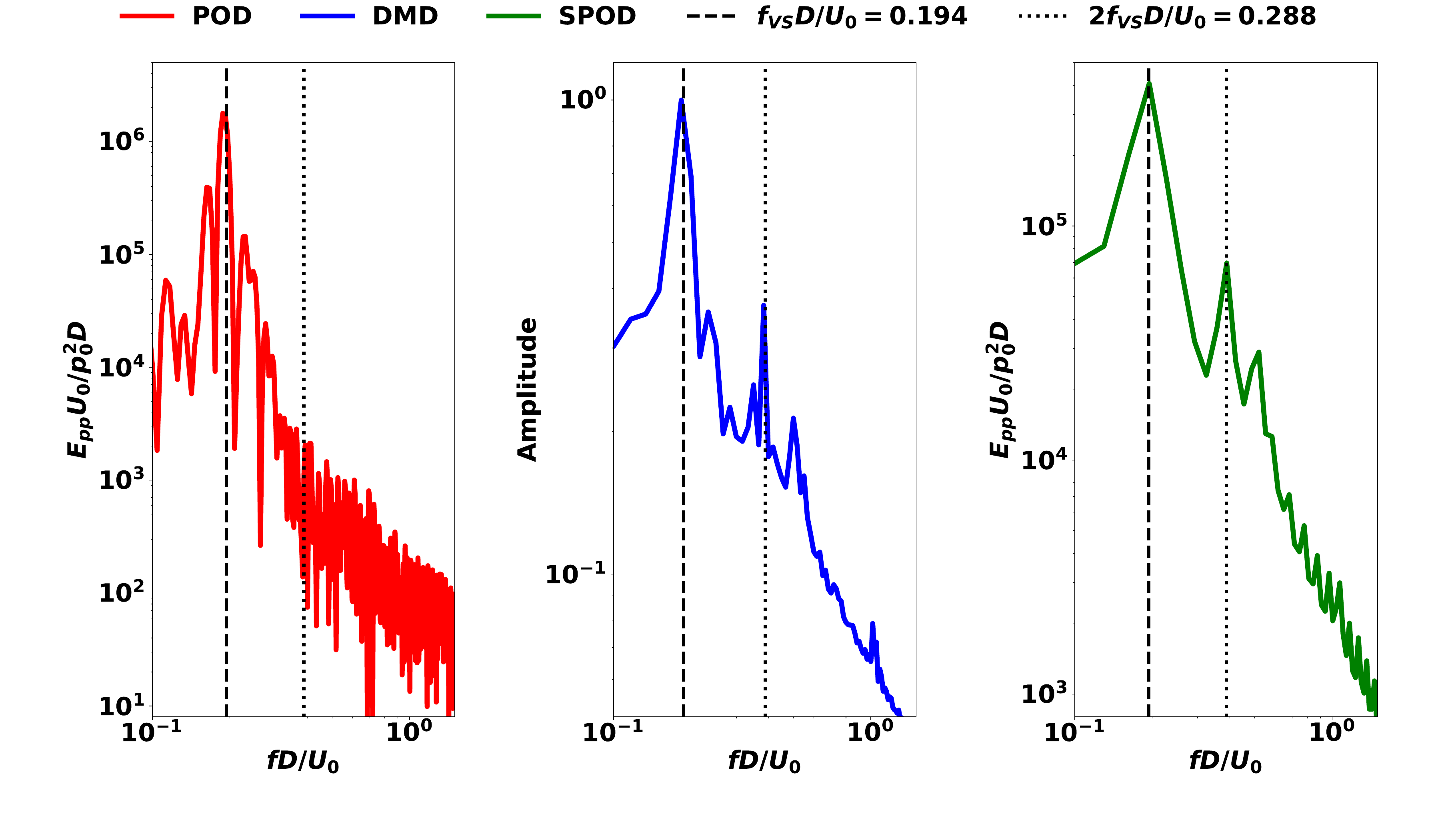}
        \label{fig:speccyl}
        }
        \caption{Comparison of the most energetic spatial correlation identified by the POD, DMD and SPOD (a), and the spectrum of the temporal coefficient of the first POD mode with the global spectra of DMD and SPOD (b) for the pressure of the cylinder at Mach number $M = 0.5$ and Reynolds number $Re_D = 1\times 10 ^4$ analyzed by Eiximeno et al. \cite{eiximeno2023hybrid} }
        \label{fig:val_cyl}
\end{figure}

As a last example, \autoref{fig:modecyl} shows the comparison of the most relevant spatial correlation of the pressure fluctuations in the wake of a cylinder at $M=0.5$ and $Re_D=1\times 10^4$ \cite{eiximeno2023hybrid} and \autoref{fig:speccyl} compares the spectrum of the temporal coefficient of this mode with the global spectra of the DMD and SPOD analysis. All three decompositions yield a similar spatial correlation for the most energetic mode and identify the same dominant frequency $fD/U_{\infty} = 0.194$ for it. Moreover, the global spectra of the DMD and SPOD also show that there is a relevant mode at the first harmonic of this frequency, $fD/U_{\infty}=0.288$. This spatial correlation and dominant frequency are coherent with the vortex shedding frequency of this case \cite{eiximeno2023hybrid}

% Results folders on MN for scalability
% Channel 180:
% /gpfs/scratch/bsc21/bsc21828/Channel180/ROMs/strong_scalability
% /gpfs/scratch/bsc21/bsc21828/Channel180/ROMs/weak_scalability
% Airplane:
% /gpfs/scratch/bsc21/bsc21828/plane_m19-290/ROMs/scalability
% Difuser:
% /gpfs/archive/hpc/bsc21/bsc21828/HiFiTurb/TC05_Standford_Diffuser/ROMs
\section{Code performance}
\label{sec:performance}
This section aims to provide a detailed study of the performance of pyLOM by analyzing its strong and weak scalability. The tests were conducted on Marenostrum 4, a supercomputer with 3456 nodes composed of two Skylake generation Intel Xeon Platinum 8160  processors. Each socket has 24 cores, resulting in a total of 48 cores per node. Every node has 96GB of RAM memory and the nodes are interconnected with an Intel Omni-Path high-performance network of 100 Gbit/s.

All the performance studies in this paper have been done using the flow dataset generated in Goc et al. \cite{goc_lehmkuhl_park_bose_moin_2021} of wall-modeled large eddy simulations of the Japanese Exploration Agency (JAXA) Standard Model (JSM) aircraft. The simulations were performed on a grid of 58.3 million points and 56 snapshots were collected.

The first analysis for the performance assessment is a strong scaling test to find the speedup of the code ($ Speedup = t(1)/t(P)$) when the number of processors is increased while the size of the problem remains constant. Ideally, a problem would scale linearly by a factor of the number of processors used, $P$. In reality, the speedup is limited by the fraction of the serial part of the software and the communication time between processors. According to Amdahl's law \cite{amdahl}, the serial part that limits the speedup in a fixed-size problem is defined by 
\begin{equation}
    \centering
    Speedup = \frac{1}{S + \frac{1-S}{P}},
    \label{eqn:amdahl}
\end{equation}
where $S$ is the proportion of the code executed in serial.

The strong scaling tests of pyLOM have been done by decomposing the whole dataset resulting in a matrix of size of $\left(5.83\times 10^7,\text{ }56\right)$. They have been run using between 8 and 84 computing nodes as 8 nodes were the minimum to fit the whole array into memory.

\begin{figure}
\centering
\includegraphics[width=0.7\textwidth]{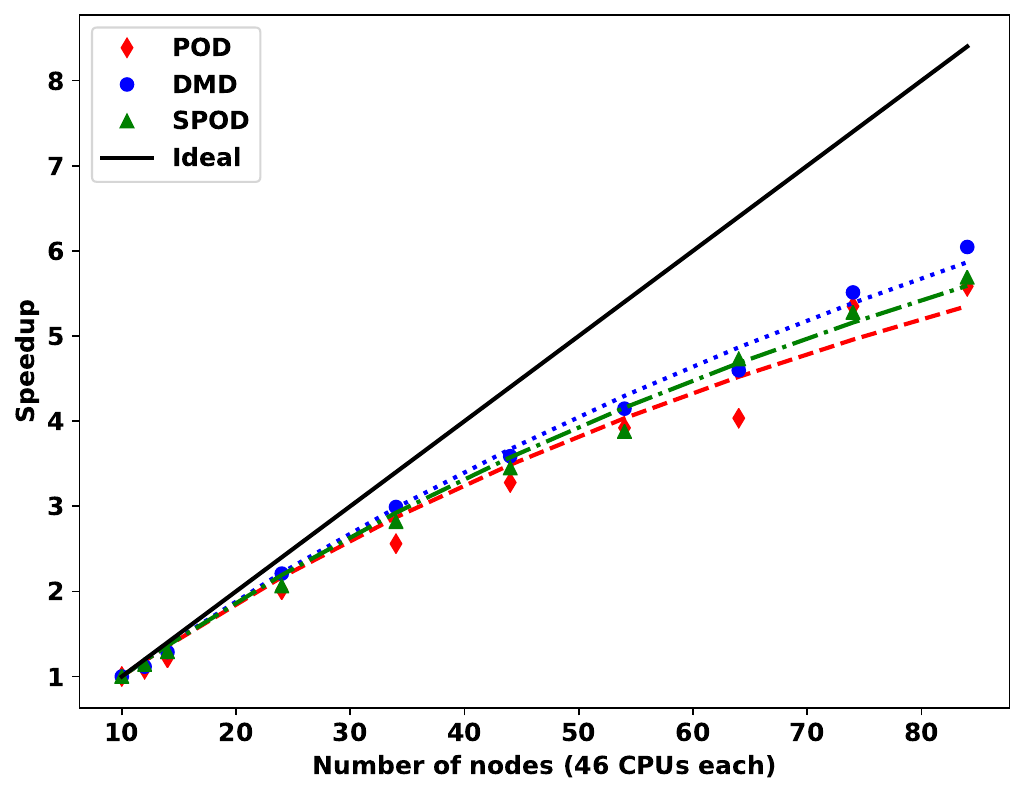}
\caption{Amdahl fit (\autoref{eqn:amdahl}) for the POD (serial part percentage of 7.70\%), DMD (serial part percentage of 5.84\%) and SPOD (serial part percentage of 6.80\%).}
\label{fig:amdahl}
\end{figure}

\autoref{fig:amdahl} shows the speedup of the three algorithms together with their corresponding fit (\autoref{eqn:amdahl}). The results show that DMD is the algorithm with the best strong scalability as only a $5.84\%$ is executed in serial, followed by SPOD with a $6.80\%$ and POD with a $7.70\%$.

In addition, two weak scaling tests, one on the number of rows ($M$) and another on the number of columns ($N$), have been performed to analyze how the algorithms scale with the amount of resources. In this case, the speedup is computed as a scaled speedup and is defined as:

\begin{equation}
    \centering
    Scaled\text{ }Speedup = \frac{t(1)}{t(P)}P
\end{equation}

For this type of problem, Gustafson \cite{gustafson} reformulated Amdahl's law by postulating that the serial part does not increase with the size of the problem and the parallel part scales linearly with the number of cores:

\begin{equation}
    \centering
    Scaled\text{ }Speedup = S + (1-S) \times P
    \label{eqn:gustafson}
\end{equation}

Weak scaling also gives a magnitude of the software efficiency, 
\begin{equation}
    \centering
    Efficiency = \frac{t(1)}{t(N_{procs})},
\end{equation}
ideally, if the workload per processor remains constant, the wall-clock execution time should be the same regardless of the number of processors. 

An important part of all three algorithms is based on the parallel QR factorization from Demmel et al. \cite{demmel_communication-optimal_2012} and according to \autoref{eqn:flops} the workload per processor depends linearly on the number of rows. This is also true for the parallel matrix-matrix product used in DMD (\autoref{eqn:flopsmatmul}) and agrees with all the operations done in serial to the snapshots matrix, such as the Fast Fourier Transform in SPOD or the mode computation in DMD. Thus, in the study of the weak scaling with the number of rows, the global matrix will increase its size as $\left(M_iP,N\right)$, where $M_i$ is the constant number of rows per processor set to $M_i=1\times 10^6$. In this test, the number of columns is set to $N=56$.

\begin{figure}
    \begin{center}
        \subfloat[]{
        \includegraphics[width = 0.485\textwidth]{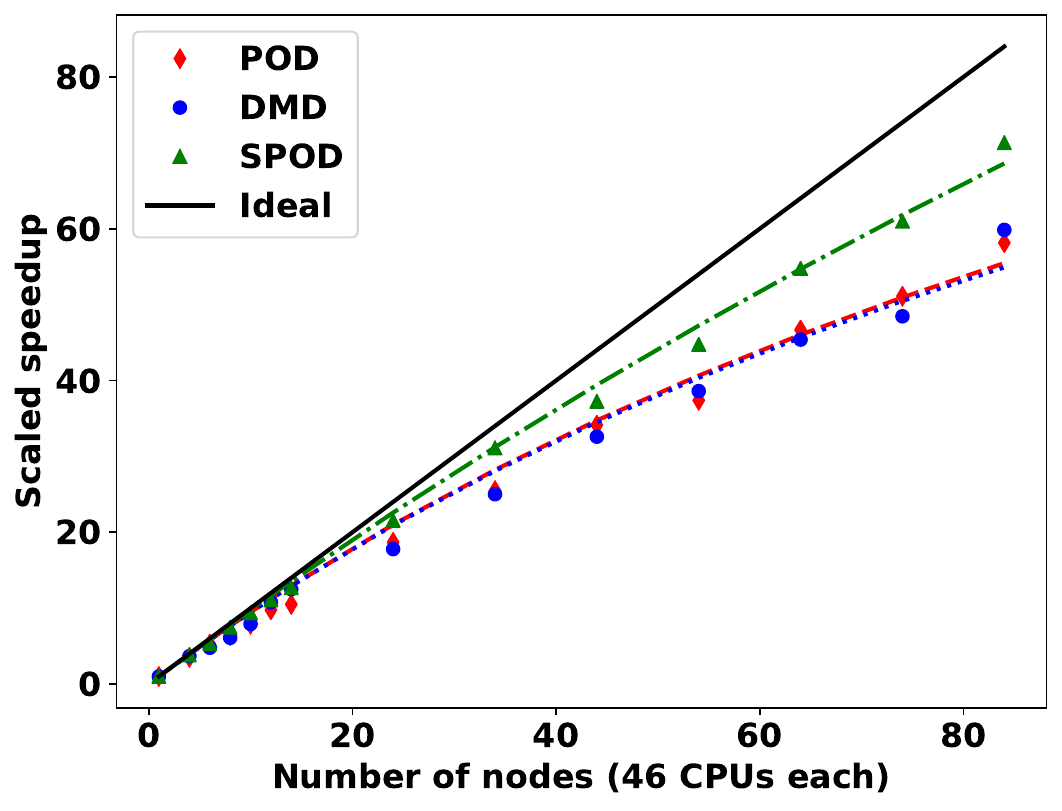}
        \label{fig:gustafson}
        }
        \subfloat[]{
        \includegraphics[width = 0.5\textwidth]{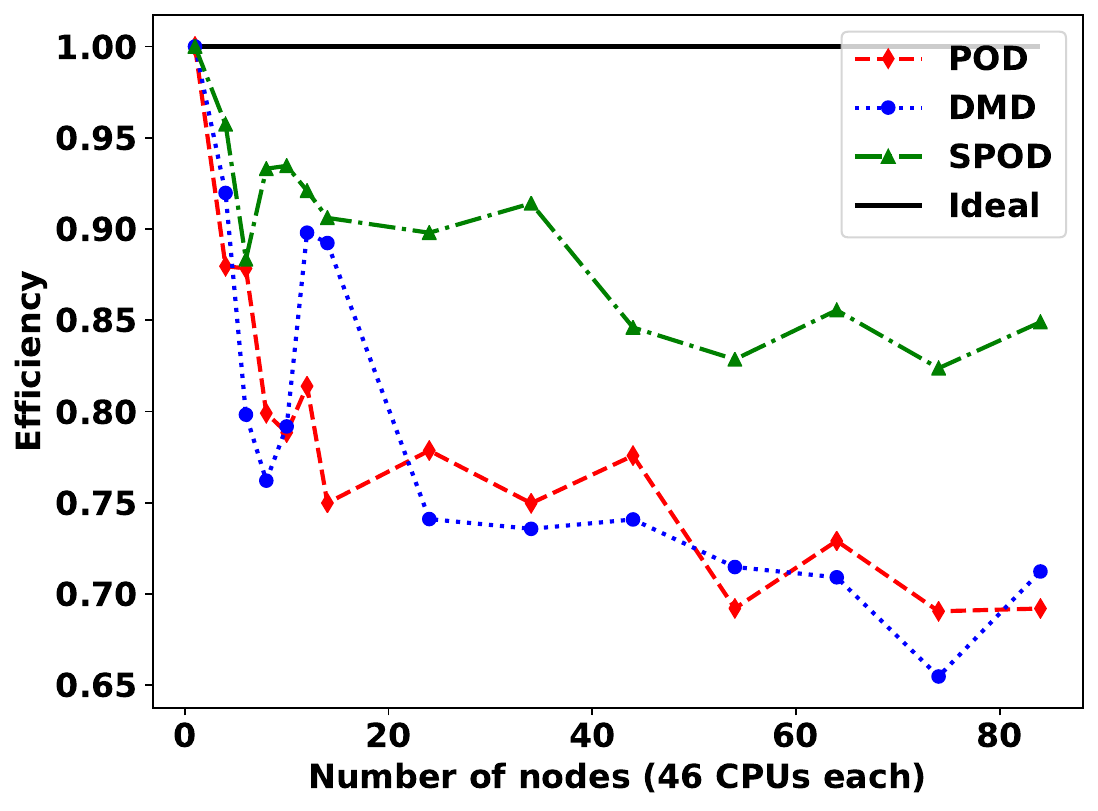}
        \label{fig:efficiency}
        }
    \end{center}
    \caption{Gustafson fit (\autoref{eqn:gustafson}) for the POD (serial part percentage of 0.62\%), DMD (serial part percentage of 0.64\%) and SPOD (serial part percentage of 0.27\%) (a) and efficiency plot for each algorithm for the weak scaling regarding the number of rows (b)}
    \label{fig:weak}
\end{figure}

\autoref{fig:gustafson} presents the scaled speedup for the three algorithms together with their fit to Gustafson's law (\autoref{eqn:gustafson}). It illustrates that SPOD is the algorithm with the best weak scaling when the number of rows scales linearly with the available resources. SPOD only has a serial fraction of 0.27\% while POD and DMD have 0.62\% and 0.64\%, respectively. Accordingly to the scaled speedup, \autoref{fig:efficiency} shows that SPOD is the algorithm with the best efficiency while POD and DMD have a  similar performance.

Despite the linear dependence of the local workload with the number of rows, it is not straightforward to find a variation of the local workload with the number of columns. The algorithms where this dependence can be estimated are the QR factorization (\autoref{alg:tsqr_svd}) and the parallel matrix-matrix product  (\autoref{alg:matmul_parañ}). In both cases, the number of operations (\autoref{eqn:flops} and \autoref{eqn:flopsmatmul}) has a strong dependence on the square of the number of columns, $N^2$.  Hence, in the weak scaling tests to $N$, the matrix increases as $\left(M/P,N_0 \sqrt{P}\right)$, where $N_0$ is the initial number of columns set to $N_0=10$. For this test, $M$ is fixed to $M=3.6\times10^7$ so that the QR factorization has the same load per processor as in the weak scaling test for $M$. This test is run with a number of nodes that results in an exact square root value to ensure an integer number of columns.

\begin{figure}
    \centering
    \includegraphics[width=0.7\textwidth]{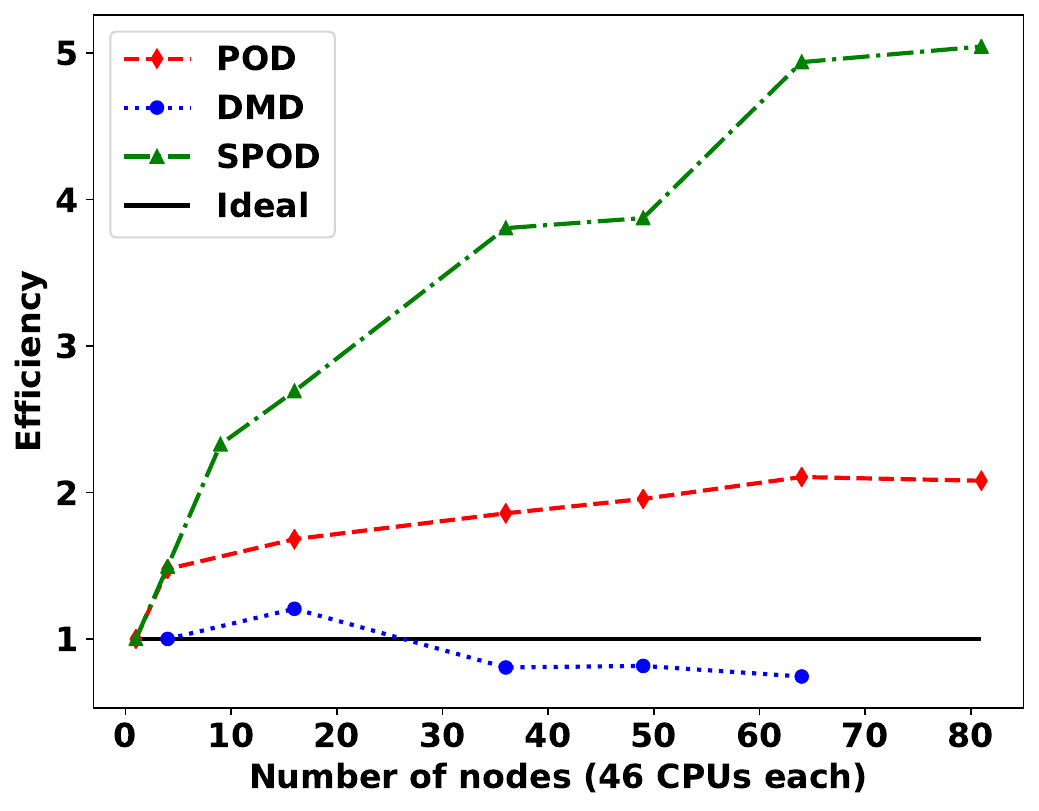}
    \caption{Efficiency of each algorithm for the weak scaling regarding the number of columns}
    \label{fig:weak2}
\end{figure}

\autoref{fig:weak2} presents the efficiency of each algorithm for the weak scaling regarding the number of columns and illustrates that in POD and SPOD it is higher than the ideal efficiency, which means that the local workload decreases with the number of processors in these two algorithms. This is in agreement with the findings reported by Demmel et al. \cite{demmel_communication-optimal_2012} on the same test, as they proved that the dependence of the QR factorization algorithm on the number of columns is not purely quadratic.  

It is important to do a profiling analysis for each algorithm to identify their limitations and bottlenecks, as well as to determine a more accurate model for the dependence of the performance with the number of columns. 

\subsection{Proper orthogonal decomposition}
\begin{figure}[h]
    \centering
    \includegraphics[width=\textwidth]{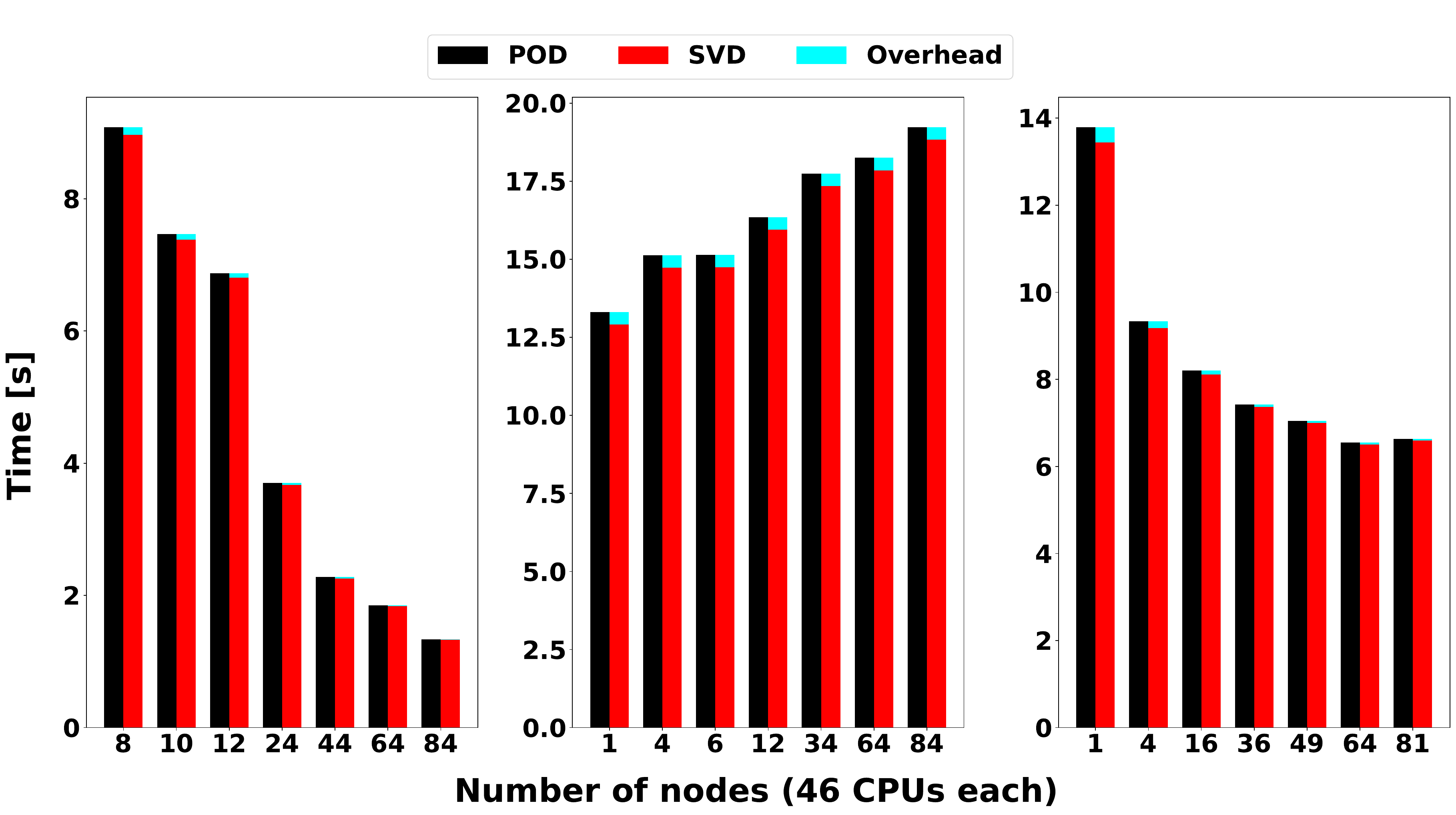}
    \caption{Profiling of the POD for the strong scaling (left), weak scaling concerning $M$ (middle) and weak scaling concerning $N$ (right).}
    \label{fig:profiling_POD}
\end{figure}

The profiling of the POD implementation for the three performance analyses conducted in this paper is shown in \autoref{fig:profiling_POD}. The SVD takes the majority of the time in all three tests, except for a small overhead that accounts for the declaration and memory allocation of the arrays. This overhead loses relevance as the local size of the problem decreases. Thus, it is reduced in the strong scaling and in the weak scaling regarding the number of columns, while it has a constant value in the weak scaling regarding the number of rows. 

The computational time of the SVD (\autoref{eqn:time2}) depends on the initial QR factorization and the number of communications. All the communicated arrays during the QR factorization are of size $\left(2N,N\right)$ with the maximum $N$ value, $N=630$, taking place for the last run of weak scaling analysis concerning the number of columns. This communication only represents 0.05\% of the available bandwidth of MareNostrum 4. Therefore, $\beta$ will be omitted from \autoref{eqn:time2}.

For a strong scaling problem, the load of the initial QR factorization diminishes with the number of cores as $2MN^2 \alpha_S/P$, where $\alpha_s$ is the time per flop during the strong scaling analysis. On the other hand, after neglecting the data exchange effects, the communication time increases as $\left(\frac{2}{3}N^3\alpha_{S} + \gamma_{S}\right)\log{P} $, where $\gamma_{S}$ is the latency of the communications.

\begin{figure}
    \begin{center}
        \subfloat[]{
            \includegraphics[width = 0.46\textwidth]{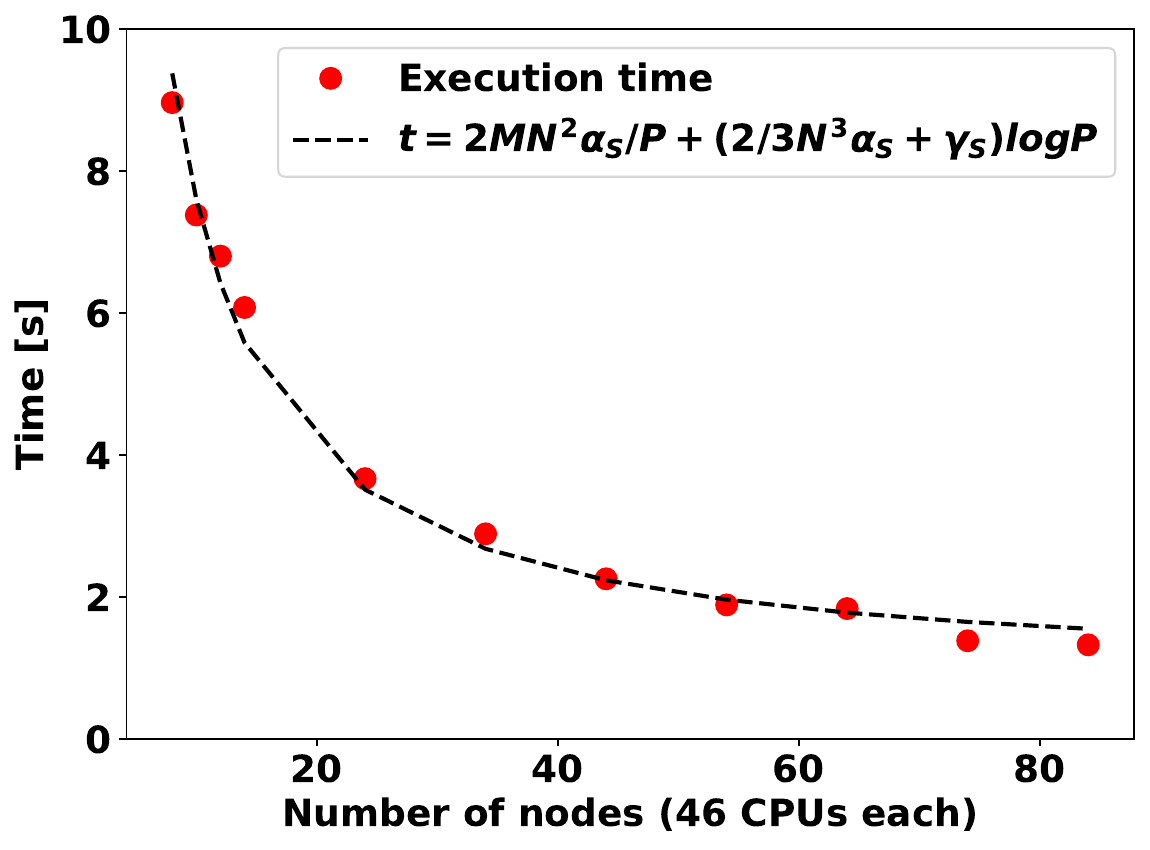}
            \label{fig:strong_fitting_SVD}
        }
        \subfloat[]{
            \includegraphics[width = 0.46\textwidth]{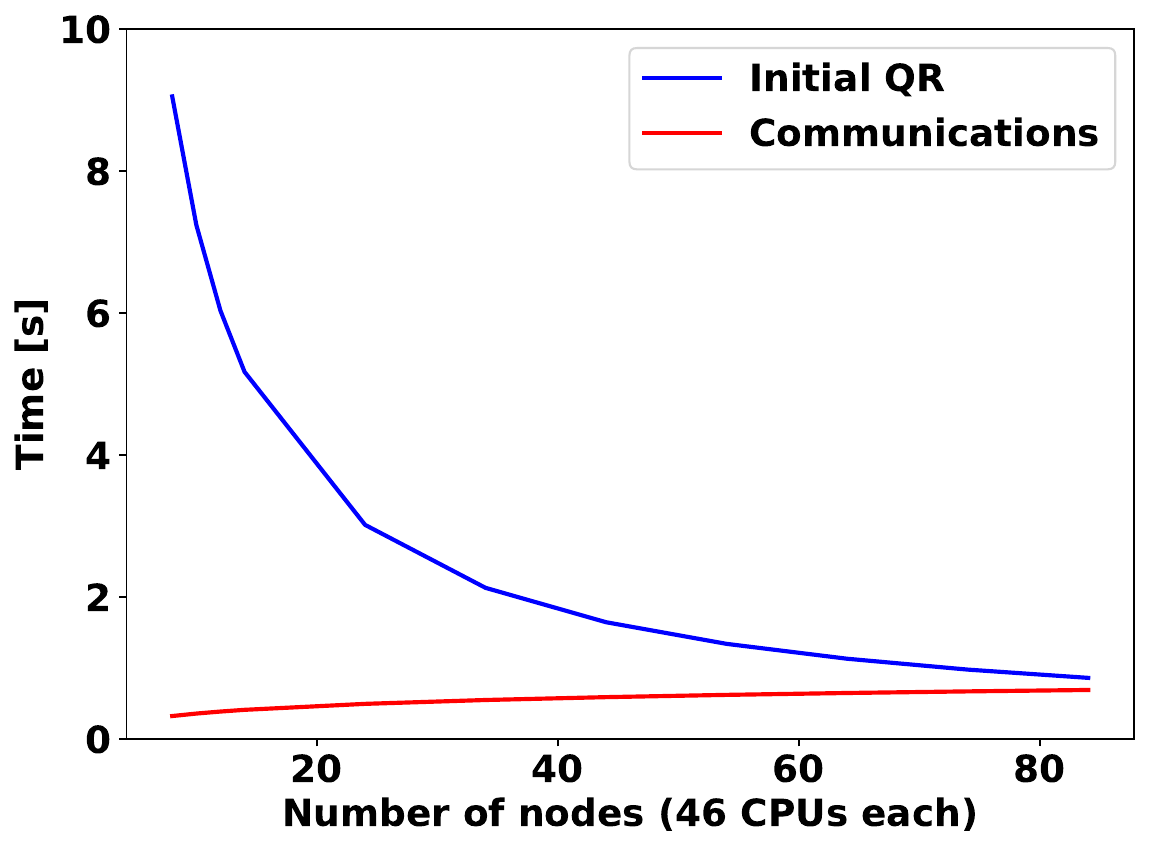}
            \label{fig:strong_profiling_SVD}
        }
    \end{center}
    \caption{(a) fit of the obtained timings for the strong scaling test of the single value decomposition to $\alpha_S=(1.109 \times 10^{-8} \pm 2.9\times 10^{-10})$ and $\gamma_S=(0.155 \pm 0.033)$, (b) comparison of the time taken by the initial QR factorization and the communications}
    \label{fig:strong_SVD}
\end{figure}

\autoref{fig:strong_fitting_SVD} shows the fit of the computing times of the single value decomposition in the strong scaling test where $\alpha_S=(1.109 \times 10^{-8} \pm 2.9\times 10^{-10})$ and $\gamma_S=(0.155 \pm 0.033)$. Despite the time per flop being 7 orders of magnitude lower than the latency, the contribution of the initial QR decomposition can not be neglected in tall and skinny matrices with a large number of rows. In fact \autoref{fig:strong_profiling_SVD} elucidates that the initial QR decomposition is the dominant factor in the computational time and that the logarithmic increase of the communications time is only relevant for the cases using a larger number of nodes. This is the reason why the execution time plotted in \autoref{fig:strong_fitting_SVD} always decays and the growth of the communication time cannot be appreciated.

Concerning the weak scaling on the number of rows, the QR factorization time is constant to $2M_iN^2\alpha_{WM}$ and the communication time increases as $\left(\frac{2}{3}N^3\alpha_{WM} + \gamma_{WM}\right) \log{P}$

\begin{figure}
    \begin{center}
        \subfloat[]{
            \includegraphics[width = 0.46\textwidth]{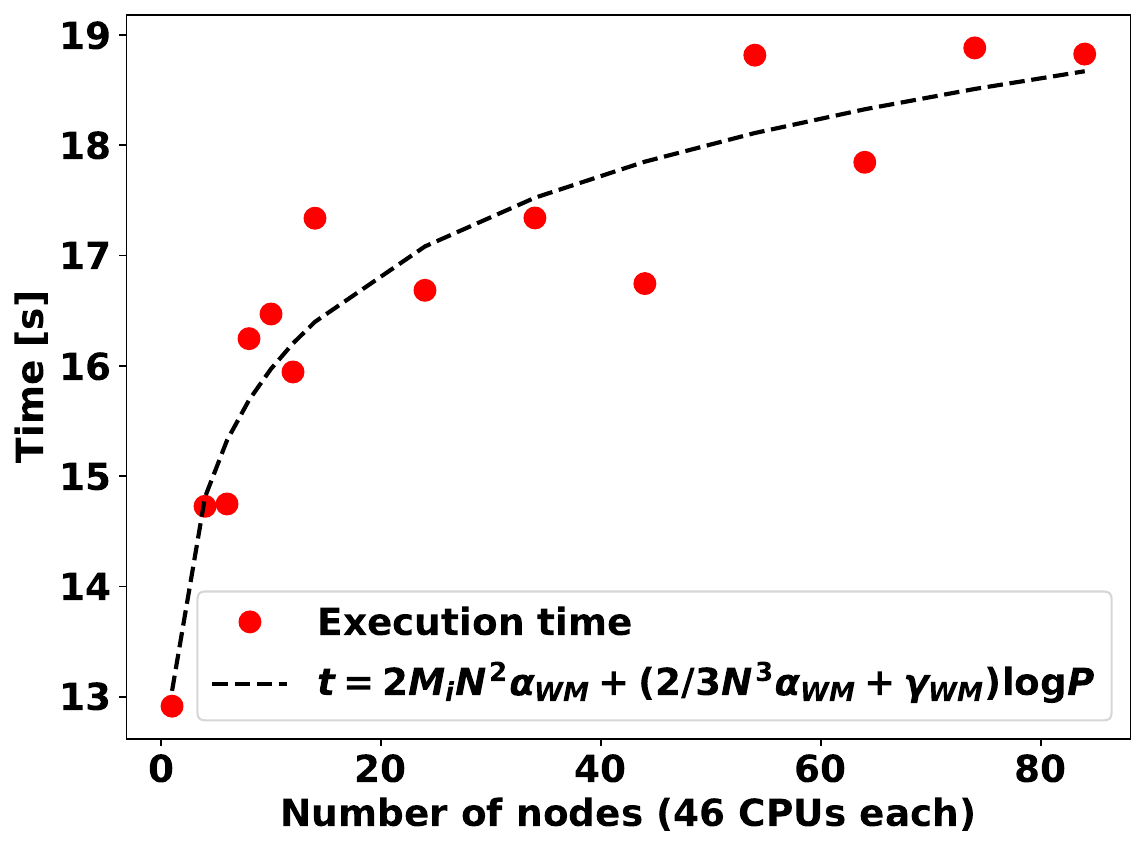}
            \label{fig:weak_fitting_SVD}
        }
        \subfloat[]{
            \includegraphics[width = 0.46\textwidth]{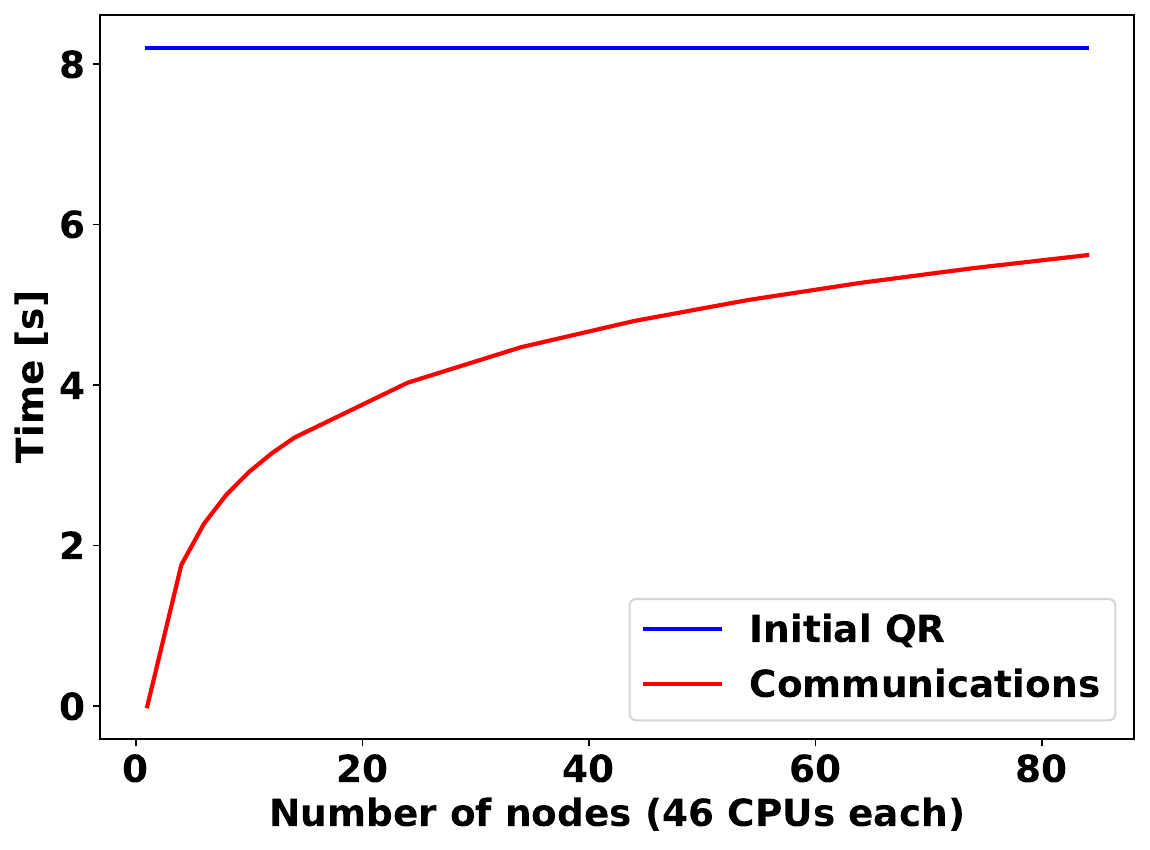}
            \label{fig:weak_profiling_SVD}
        }
    \end{center}
    \caption{(a) fit of the obtained timings for the weak scaling test concerning the number of columns to $\alpha_{WM}=(5.33 \times 10^{-10} \pm 2.14\times 10^{-10})$ and $\gamma_{WM}=(1.268 \pm 0.127)$, (b) comparison of the time taken by the initial QR factorization and the communications}
    \label{fig:weak_SVD}
\end{figure}

Analogously, $\alpha_{WM}$ and $\gamma_{WM}$ are the time per flop and the latency for the weak scaling test regarding the number of rows. \autoref{fig:weak_fitting_SVD} shows the fit of \autoref{eqn:time2} with $\alpha_{WM}=(5.33 \times 10^{-10} \pm 2.14\times 10^{-10})$ and $\gamma_{WM}=(1.268 \pm 0.127)$. Although the difference between $\alpha_{WM}$ and $\gamma_{WM}$ is even bigger than for the strong scaling test, \autoref{fig:weak_profiling_SVD} indicates that the initial QR factorization is still the dominant factor in the computational time. Moreover, \autoref{fig:weak_profiling_SVD} elucidates that the initial QR factorization time does not depend on the number of processors and that the time increase detected in \autoref{fig:weak_fitting_SVD} is due to the rise of the communications costs with the logarithm of the number of processors, $\log{P}$.

\begin{figure}
    \begin{center}
        \subfloat[]{
        \includegraphics[width = 0.46\textwidth]{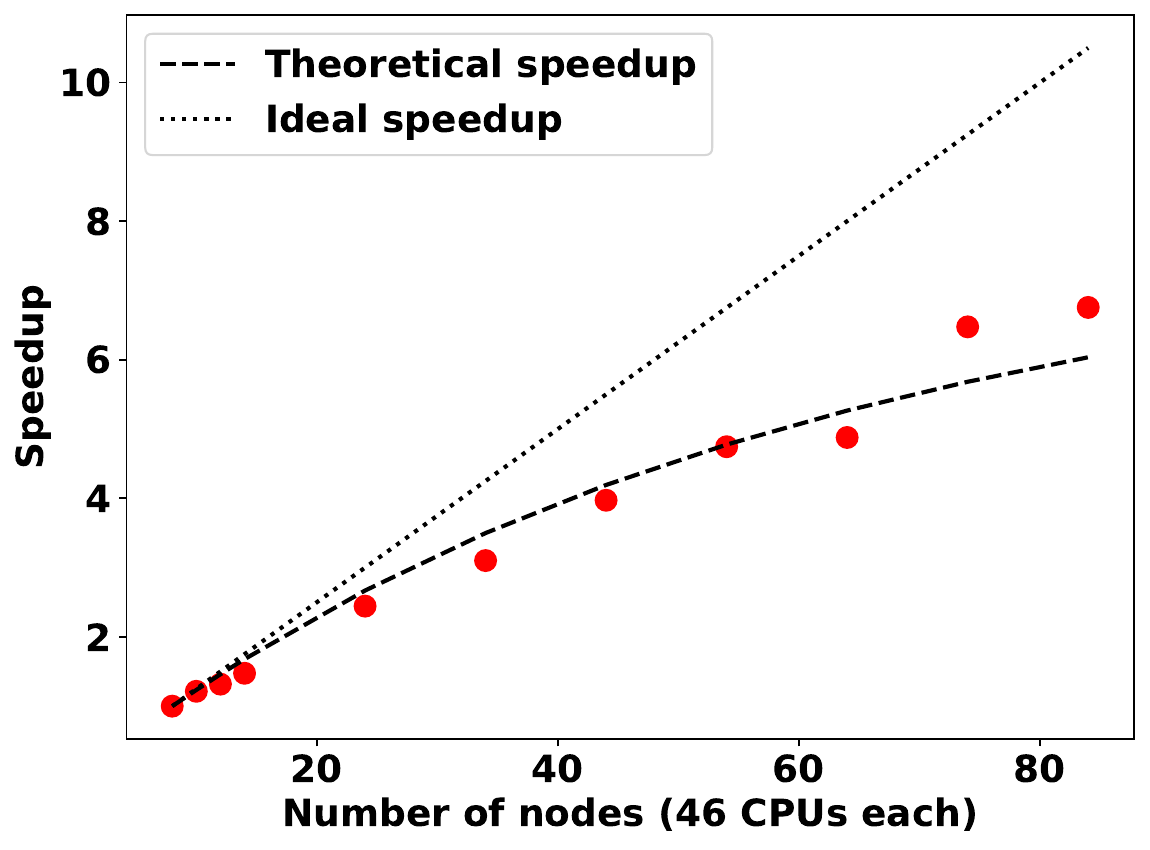}
        \label{fig:speedup_SVD}
        }
        \subfloat[]{
        \includegraphics[width = 0.46\textwidth]{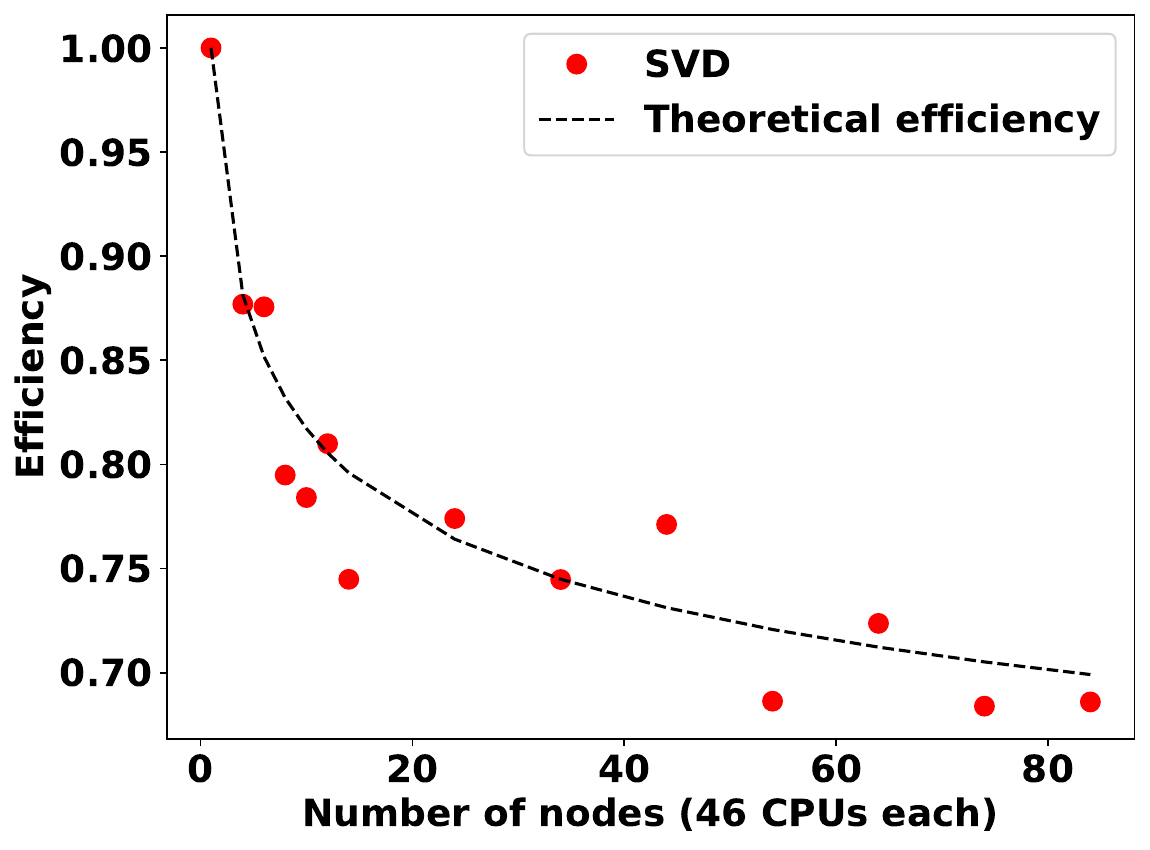}
        \label{fig:weak_efficiency_SVD}
        }
    \end{center}
    \caption{Comparison between the theoretical and ideal speedup (a) and efficiency (b) of the SVD}
    \label{fig:limits_SVD}
\end{figure}

The communications time increment is the sole responsible for the decay in the speedup and efficiency of the algorithms. \autoref{fig:limits_SVD} plots the comparison between the ideal speedup and efficiencies, their theoretical values considering the communications cost and the results obtained in the computations. This figure shows that the obtained times are aligned with the expected performance.

The results of \autoref{fig:weak2} identify that the dependence of the load on the number of columns is not purely quadratic. Hence, an extra dependence must be found from the obtained timings. The cost of the QR decomposition in this case is
\begin{equation}
    \centering
    \frac{2M\left(N_0\sqrt{P}\right)^x}{P}\alpha_{WN}
    \label{eqn:cweakn}
\end{equation}
and the time for the communications
\begin{equation}
    \centering
    \left(\frac{2}{3}N_0^3P\sqrt{P}\alpha_{WN} + \gamma_{WN}\right)\log{P}
    \label{eqn:kweakn}
\end{equation}

As the load is equivalent to the one used in the weak scaling test concerning the number of rows, the time per flop, $\alpha_{WN}$, can be taken as the mean value of the time per flop obtained in the fit of \autoref{fig:weak_fitting_SVD} ($\alpha_{WN} = \overline{\alpha}_{WM} = 5.33 \times 10^{-10}$). Then the correct exponent for the number of columns is $x = 1.38$ instead of $x=2$.

\begin{figure}
    \centering
    \includegraphics[width=0.7\textwidth]{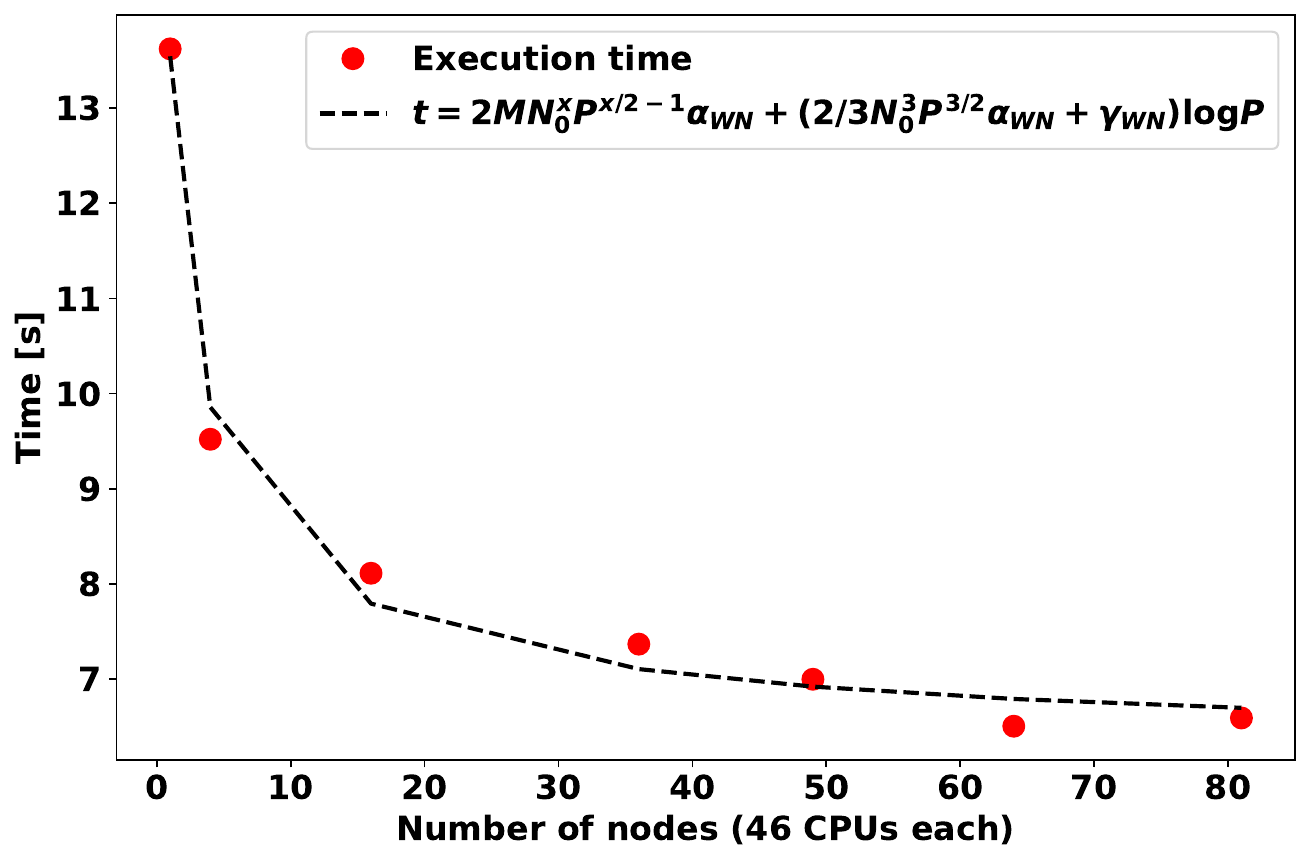}
    \caption{Fit of the obtained timings to find $x=(1.387 \pm 0.005)$ and $\gamma_{WN}=(0.723 \pm 0.049)$}
    \label{fig:weak2_fitting_SVD}
\end{figure}

\subsection{Dynamic mode decomposition}
\autoref{fig:profiling_DMD} shows the profiling of each function in DMD for each of the performance tests. In this case, the overhead not only includes the arrays declaration and memory allocation but also the splitting of the snapshots in $Y_1$ (\autoref{eqn:y1}) and $Y_2$ (\autoref{eqn:y2}) and the truncation of the SVD results. As in POD, the overhead reduces with the number of rows per processor as the arrays to allocate are smaller. Moreover, the time to truncate and to split in $Y_1$ and $Y_2$ also diminishes with the local number of rows.

\begin{figure}[b!]
    \centering
    % First part: Image
    \includegraphics[width=\textwidth]{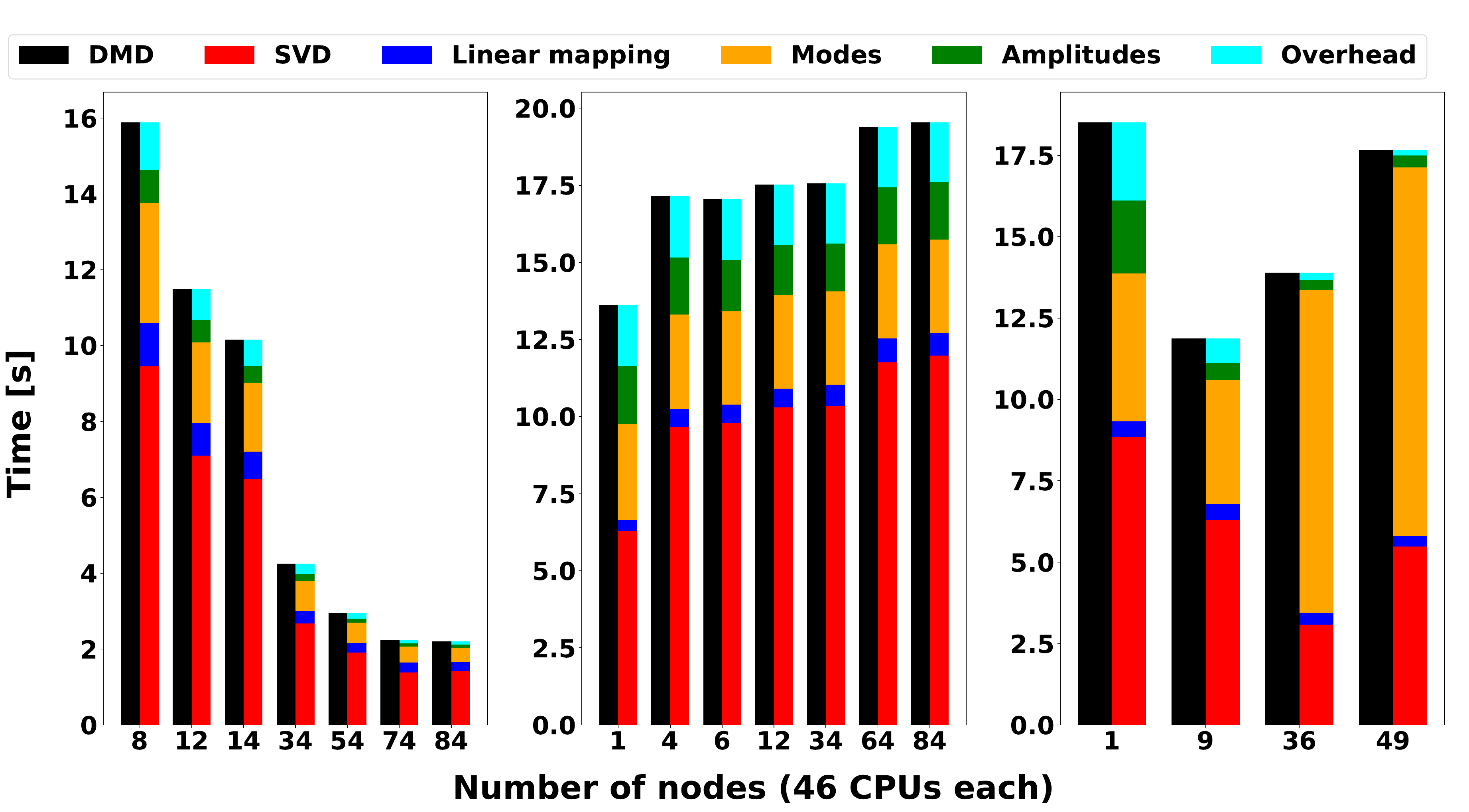}
    \caption{Profiling of the DMD for the strong scaling (left), weak scaling concerning \( M \) (middle) and weak scaling concerning \( N \) (right)}
    \label{fig:profiling_DMD}
\end{figure}
\begin{figure}\ContinuedFloat    
    % Second part: First table
    \begin{minipage}{\textwidth}
        \centering
        \begin{tabularx}{\textwidth}{p{3.3cm}XXXXXXX}
\hline
 & \textbf{8} & \textbf{12} & \textbf{14} & \textbf{34} & \textbf{54} & \textbf{74} & \textbf{84} \\
\hline
\textbf{SVD} & 59.5 & 61.8 & 63.9 & 63.0 & 64.9 & 61.8 & 64.9 \\
\textbf{Linear mapping} & 7.3 & 7.5 & 7.0 & 7.6 & 8.5 & 11.7 & 10.6 \\
\textbf{Modes} & 19.8 & 18.4 & 17.9 & 18.6 & 18.0 & 18.6 & 17.2 \\
\textbf{Amplitudes} & 5.5 & 5.2 & 4.3 & 4.3 & 3.8 & 3.7 & 3.6 \\
\textbf{Overhead} & 7.9 & 7.1 & 6.8 & 6.5 & 4.8 & 4.2 & 3.7 \\
\hline
\end{tabularx}

        \captionof{table}{Percentage of execution time taken by each function for several numbers of nodes for the strong scaling in the DMD}
        \label{tab:strong_DMD}
    \end{minipage}

    \vspace{0.3cm} % Space between first and second table

    % Third part: Second table
    \begin{minipage}{\textwidth}
        \centering
        \begin{tabularx}{\textwidth}{p{3.3cm}XXXXXXX}
\hline
 & \textbf{1} & \textbf{4} & \textbf{6} & \textbf{12} & \textbf{34} & \textbf{64} & \textbf{84} \\
\hline
\textbf{SVD} & 46.2 & 56.4 & 57.4 & 58.8 & 58.8 & 60.7 & 61.3 \\
\textbf{Linear mapping} & 2.6 & 3.4 & 3.5 & 3.5 & 4.0 & 4.0 & 3.7 \\
\textbf{Modes} & 22.8 & 17.8 & 17.8 & 17.3 & 17.3 & 15.7 & 15.5 \\
\textbf{Amplitudes} & 13.8 & 10.8 & 9.7 & 9.3 & 8.8 & 9.5 & 9.5 \\
\textbf{Overhead} & 14.5 & 11.6 & 11.6 & 11.2 & 11.1 & 10.0 & 9.9 \\
\hline
\end{tabularx}

        \captionof{table}{Percentage of execution time taken by each function for several numbers of nodes for the weak scaling concerning the number of rows in the DMD}
        \label{tab:weak_DMD}
    \end{minipage}
    \vspace{0.3cm}
    % Fourth part: Third table
    \begin{minipage}{\textwidth}
        \centering
        \begin{tabularx}{\textwidth}{p{3.3cm}XXXX}
\hline
 & \textbf{1} & \textbf{9} & \textbf{36} & \textbf{49} \\
\hline
\textbf{SVD} & 47.8 & 53.1 & 22.2 & 31.0 \\
\textbf{Linear mapping} & 2.6 & 4.1 & 2.6 & 1.9 \\
\textbf{Modes} & 24.5 & 31.9 & 71.3 & 64.0 \\
\textbf{Amplitudes} & 12.1 & 4.5 & 2.3 & 2.1 \\
\textbf{Overhead} & 13.0 & 6.4 & 1.6 & 1.0 \\
\hline
\end{tabularx}

        \captionof{table}{Percentage of execution time taken by each function for several numbers of nodes for the weak scaling concerning the number of columns in the DMD}
        \label{tab:weak2_DMD}
    \end{minipage}
\end{figure}

\begin{figure}
    \centering
    \includegraphics[width=\textwidth]{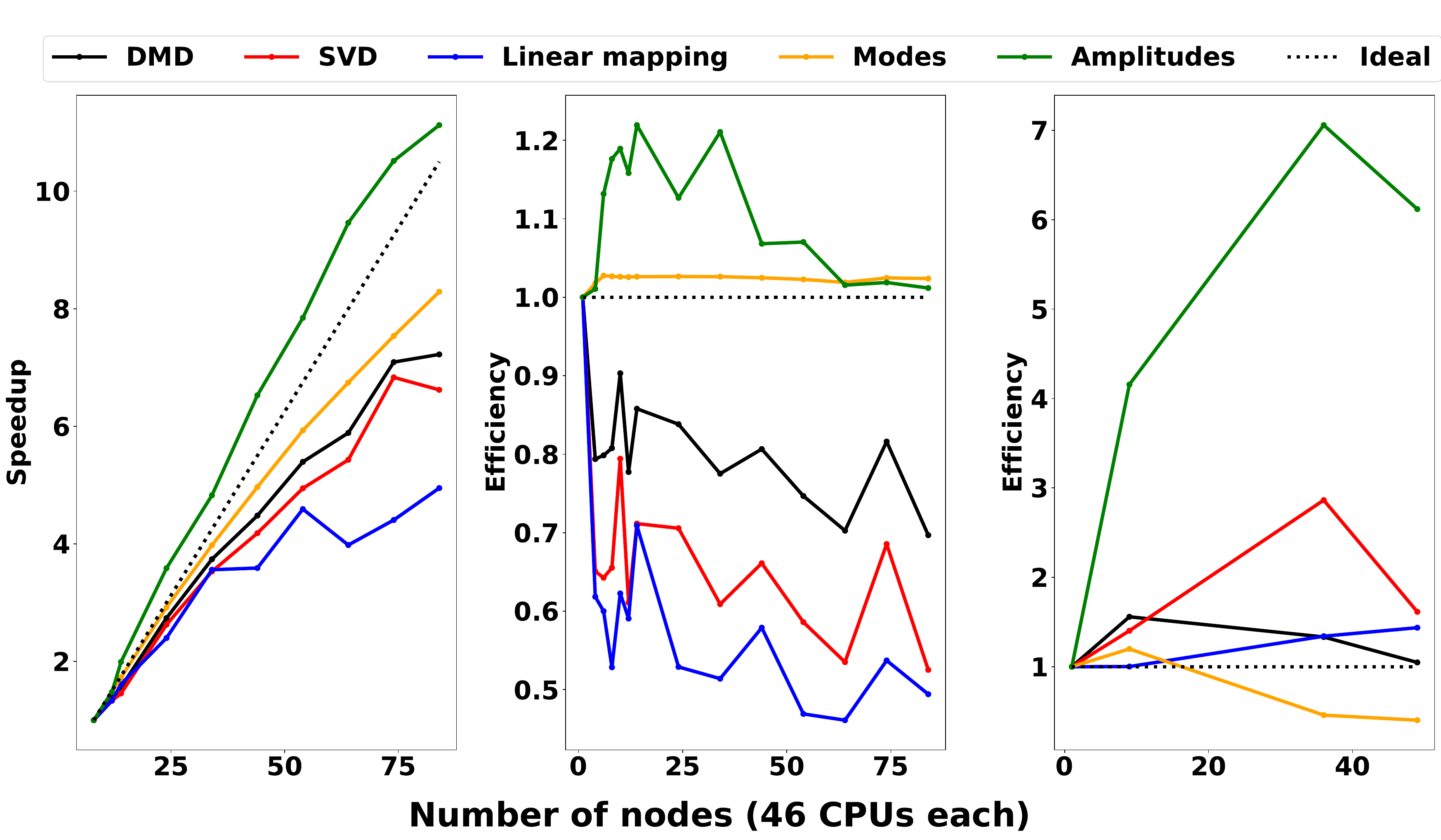}
    \caption{Speedup of the DMD for the strong scaling (left) and efficiency for weak scaling concerning $M$ (middle) and weak scaling concerning $N$ (right) }
    \label{fig:speedup_DMD}
\end{figure}

The time percentage taken by each function in the strong and weak concerning $M$ and $N$ scaling tests can be found in tables \ref{tab:strong_DMD}, \ref{tab:weak_DMD} and \ref{tab:weak2_DMD}, respectively. Except for the last runs of the weak scaling concerning the number of columns, the computing time is dominated by the single value decomposition, which takes between 59.5\% and 64.9\% of the total time in the strong scaling test and between 46.2\% and 61.3\% in the weak scaling test concerning the number of rows. It is no surprise then that the speedup and the efficiency regarding the number of rows for the DMD algorithm present a similar behavior to the ones for the POD (\autoref{fig:speedup_DMD}). However, it is important to understand the scalability of the remaining functions to get a full picture of the DMD performance.

To begin with, the linear mapping computation has limited speedup and efficiency due to the increase in the costs of the communications in the parallel matrix-matrix multiplication from \autoref{alg:matmul_parañ}. The time taken by this product (\autoref{eqn:timematmul2}) can be split between the time used for the local matrix product and the time taken for the reduction and addition of each local result. The communicated arrays are of size $(N_r,N)$. Although in these tests no truncation is considered ($N_r = N$), the data transfer will never use more than the 0.025\% of the available bandwidth and its effects can be neglected on the total time of the algorithm. Thus, all the terms containing $\beta$ in \autoref{eqn:timematmul2} are neglected and $N_r$ is considered to be $N$.

For a strong scaling test, the multiplication time decays with the number of processors as $2MN^2\alpha_S/P$ while the reduction time increases linearly with the number of processors $  \left(N^2\alpha_S + \gamma_S\right)\left(P-1\right)$

\begin{figure}
    \begin{center}
        \subfloat[]{
            \includegraphics[width = 0.46\textwidth]{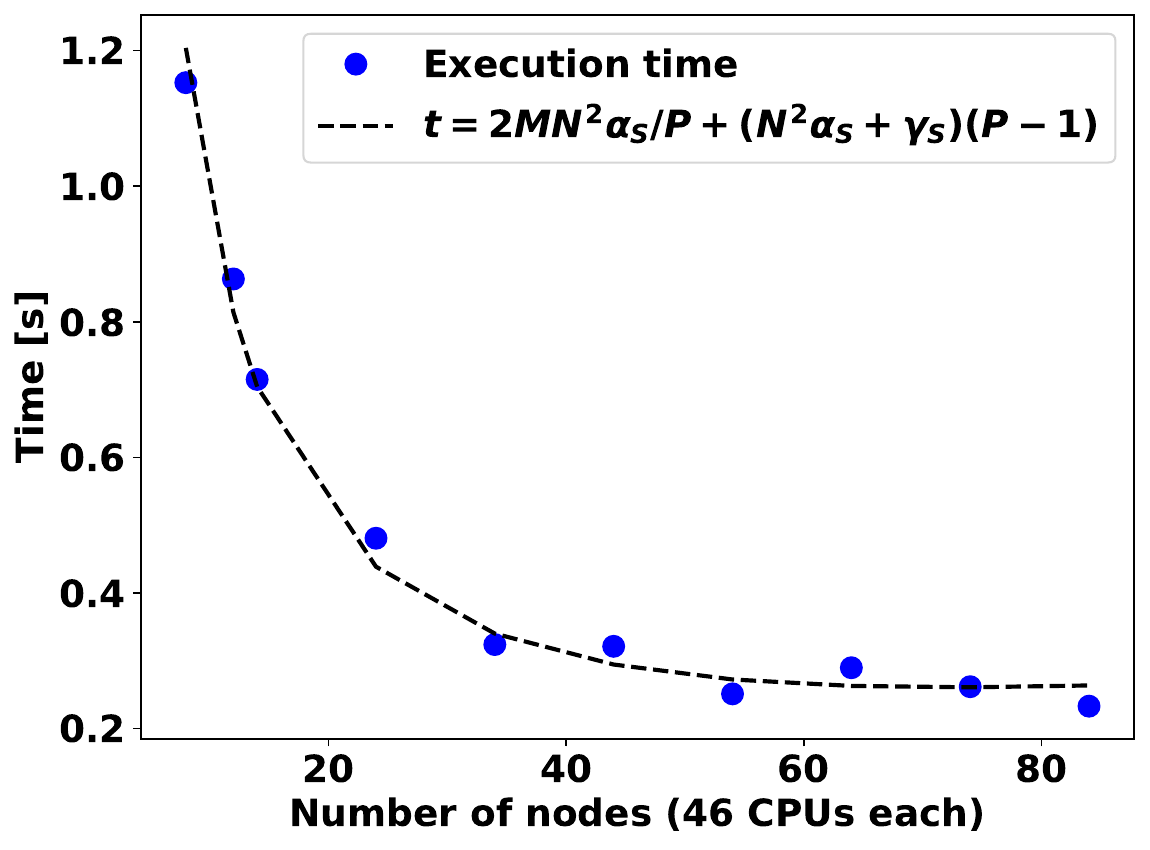}
            \label{fig:strong_fitting_linearmap}
        }
        \subfloat[]{
            \includegraphics[width = 0.46\textwidth]{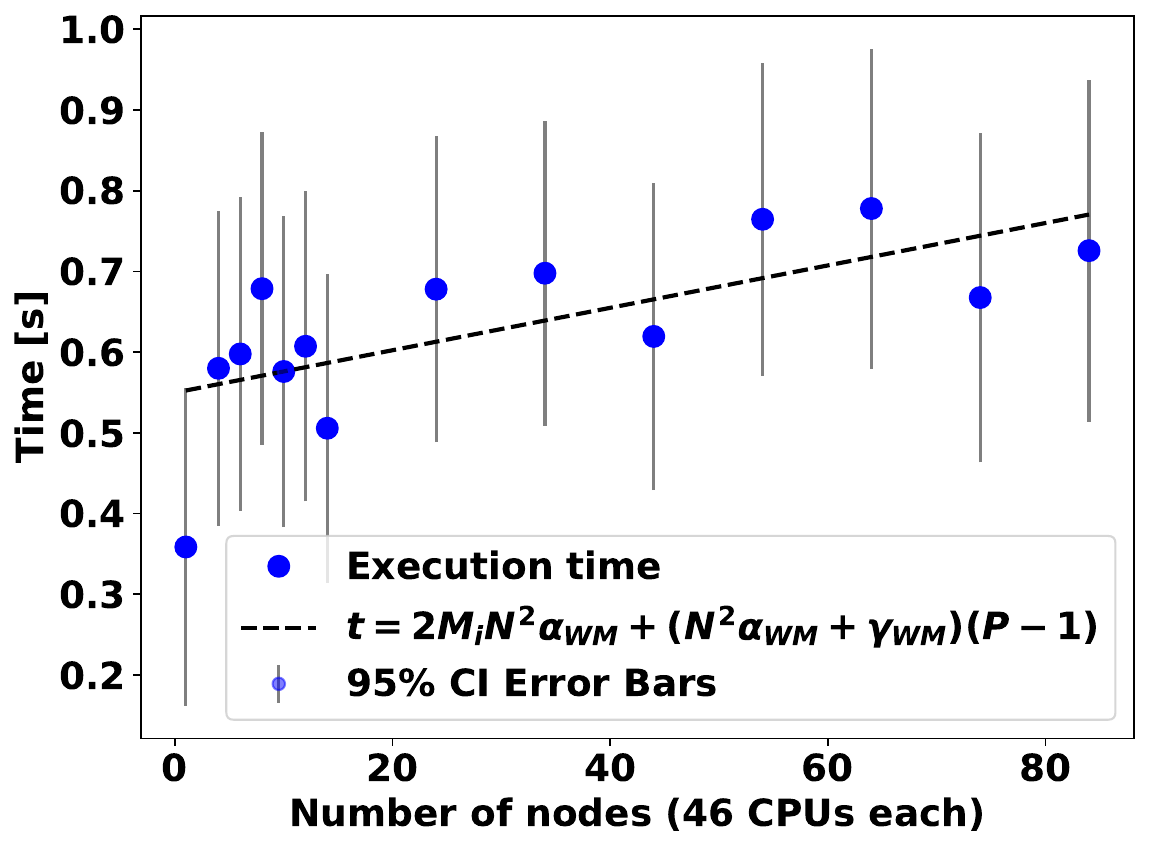}
            \label{fig:weak_fitting_linearmap}
        }
    \end{center}
    \caption{(a) fit of the obtained timings for the strong scaling test of the parallel matrix-matrix multiplication to find $\alpha_S=(1.109 \times 10^{-8} \pm 2.9\times 10^{-10})$ and $\gamma_S=(0.155 \pm 0.033)$, (b) fit of the obtained timings for the weak scaling test concerning the number of columns to find $\alpha_{WM}=(1.76 \times 10^{-10} \pm 1.1\times 10^{-11})$ and $\gamma_{WM}=(2.62 \times 10^{-3} \pm 8.2 \times 10^{-4})$ together with the 95\% confidence intervals}
    \label{fig:fitting_linearmap}
\end{figure}

\autoref{fig:strong_fitting_linearmap} shows the fit of the obtained timings for the linear mapping obtaining $\alpha_S=(1.109 \times 10^{-8} \pm 2.9\times 10^{-10})$ and $\gamma_S=(0.155 \pm 0.033)$. Although the time of the local matrix-matrix product decreases with the number of processors, there is no time reduction when running with more than 54 nodes. After this point, the communications take as much time as the multiplication. 

For the weak scaling test, the local matrix-matrix product time is constant to 
\begin{equation}
    2M_iN^2\alpha_{WM}
    \label{eqn:cweakm_matmul}
\end{equation}
and the communication time still increases linearly with the number of processors
\begin{equation}
    \left(N^2\alpha_{WM}+ \gamma_{WM}\right)\left(N-1\right)
    \label{eqn:kweakm_matmul}
\end{equation}
leading to a linear time increase in the parallel matrix-matrix product as the number of processors increases. \autoref{fig:weak_fitting_linearmap} presents the fit of the execution times obtaining $\alpha_{WM}=(1.76 \times 10^{-10} \pm 1.1\times 10^{-11})$ and $\gamma_{WM}=(2.62 \times 10^{-3} \pm 8.2 \times 10^{-4})$. The small computational time of the linear mapping makes this analysis sensitive to machine performance oscillations, in particular to variations in the latency during node-to-node communications. Those deviations yield a large mean standard error for the predicted timings with the regression and an interval in the prediction of the execution timings with 95\% confidence of 0.195 seconds.

\begin{figure}
    \centering 
    \includegraphics[width=0.7\textwidth]{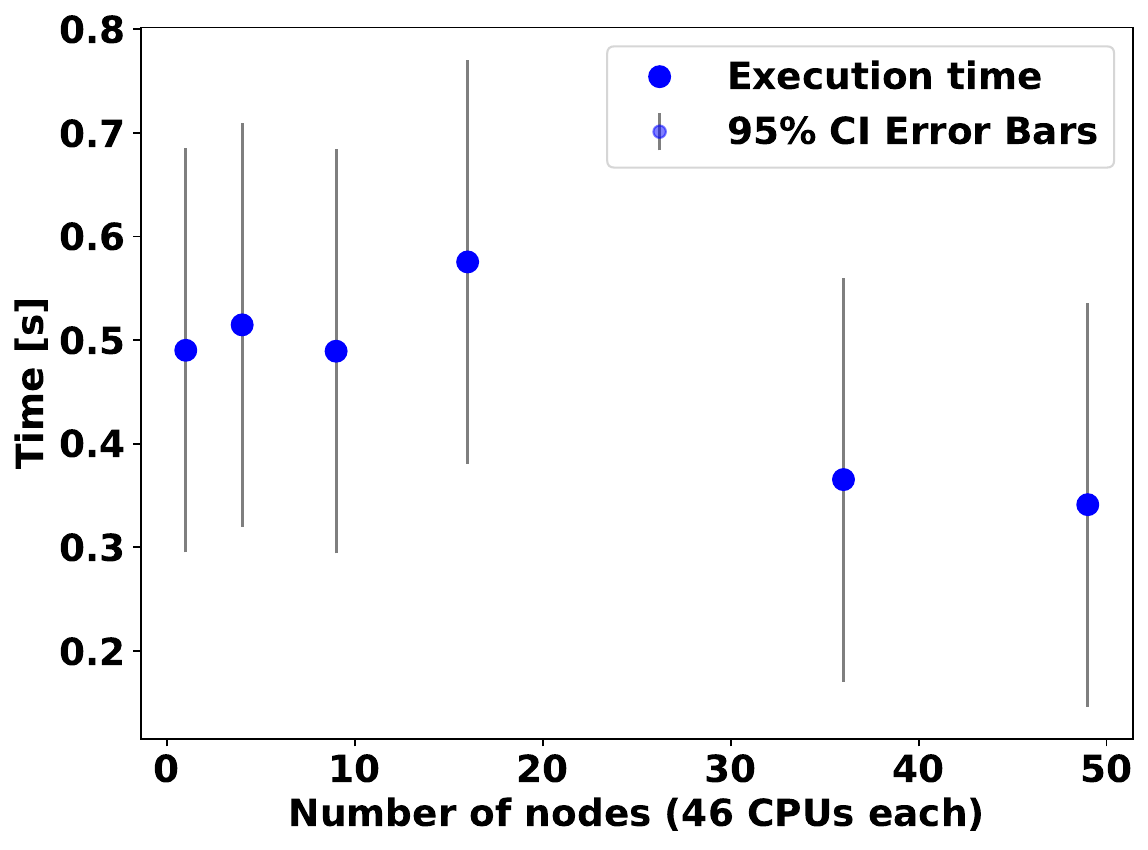}
    \caption{Execution time of the linear mapping computation according to \autoref{eqn:dmdmodes} for the weak scaling analysis regarding the number of columns}
    \label{fig:weak2_linearmap}
\end{figure}

The timings of the weak scaling regarding the number of columns for the linear mapping computation are presented in \autoref{fig:weak2_linearmap}. The reduced number of data points in this study and the sensitivity to the latency oscillations already seen in the weak scaling regarding $M$, make it difficult to identify the tendency followed by the computing time in this scenario.

\begin{figure}
    \centering 
    \includegraphics[width=\textwidth]{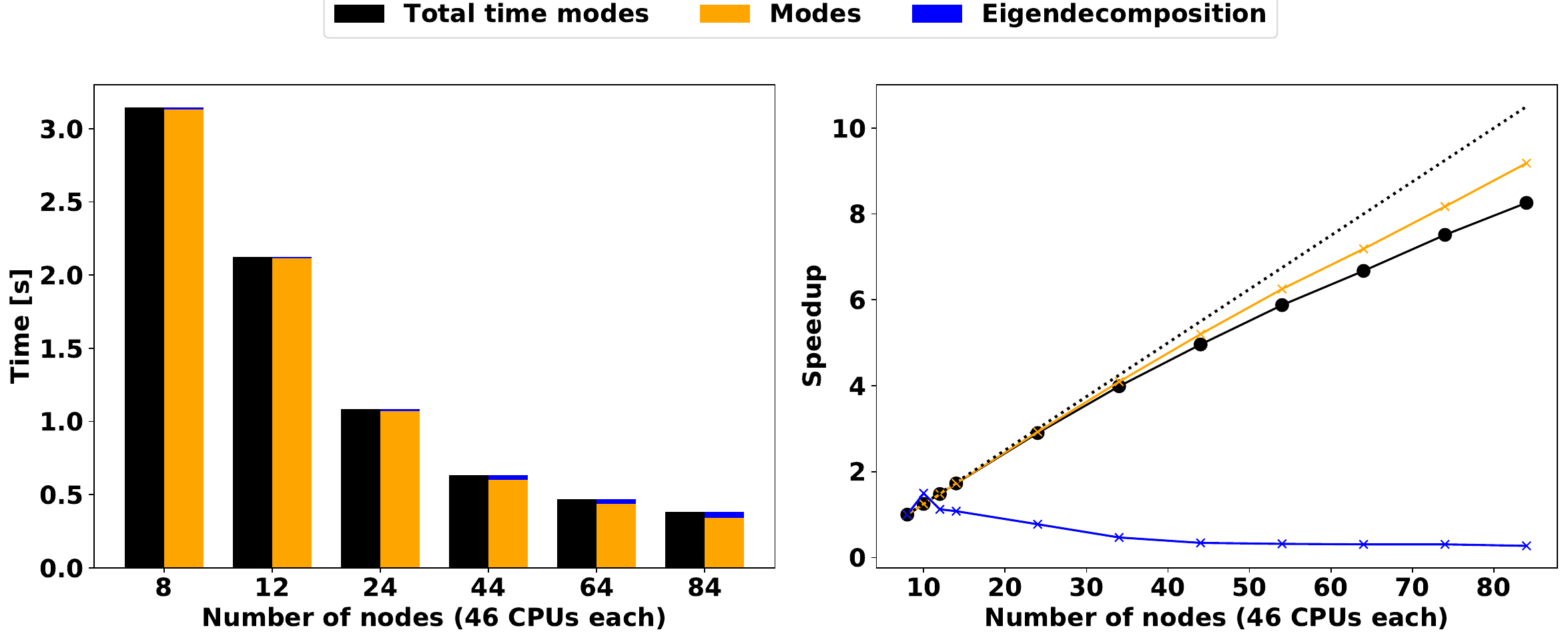}
    \caption{Profiling of the mode computation (left) and speedup of the eigendecomposition and the matrix product from \autoref{eqn:dmdmodes} (right)}
    \label{fig:speedup_modes}
\end{figure}

On the other hand, the computation of the modes presents a nearly ideal speedup. \autoref{fig:speedup_modes} shows that the origin of such an ideal performance is the perfect scalability of the serial matrix product of \autoref{eqn:dmdmodes}. The global speedup is slightly affected by the time taken by the eigendecomposition on the linear mapping because its size does not depend on the number of rows of the problem. Hence, the computing time of the eigendecomposition remains constant and its time gains relevance as the computation of the modes gets faster.

\begin{figure}
    \centering 
    \includegraphics[width=0.7\textwidth]{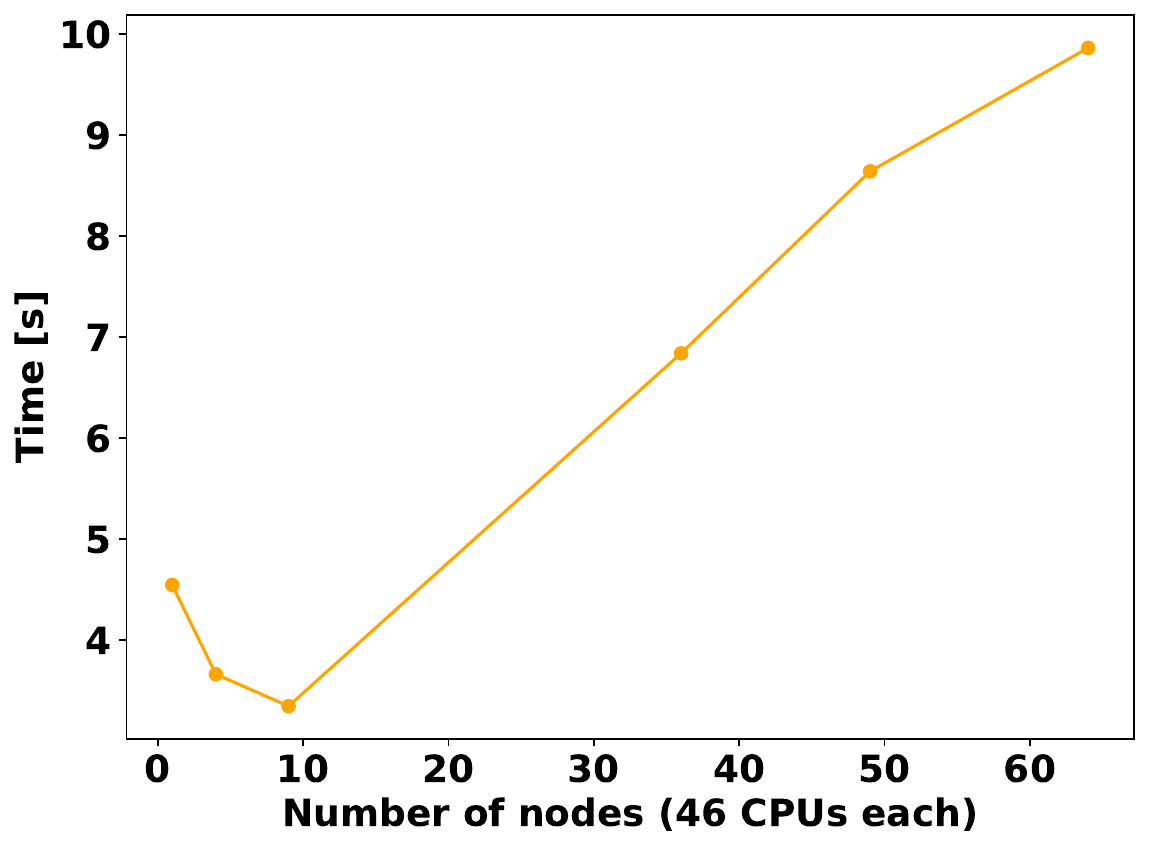}
    \caption{Execution time of the mode computation according to \autoref{eqn:dmdmodes} for the weak scaling analysis regarding the number of columns}
    \label{fig:weak2_modes}
\end{figure}

This perfect scalability is directly translated to an ideal efficiency on the weak scaling test regarding the number of rows (\autoref{fig:speedup_DMD}). The time taken by the product of \autoref{eqn:dmdmodes} is constant when the local number of rows does not change and the time used in the eigendecomposition does not depend on the number of rows of the problem. However, \autoref{fig:weak2_modes} shows that the time to compute the modes in the weak scaling regarding the number of columns increases with the number of processors as this function has a greater dependence on the number of columns than the QR factorization.

Finally, the behavior of the amplitude computation and sorting is discussed. This function stands out from \autoref{fig:speedup_DMD} because it presents hyper-speedup in the strong scaling test and efficiency over the ideal in both the weak scaling test regarding the number of rows and the number of columns.

\begin{figure}
    \centering 
    \includegraphics[width=\textwidth]{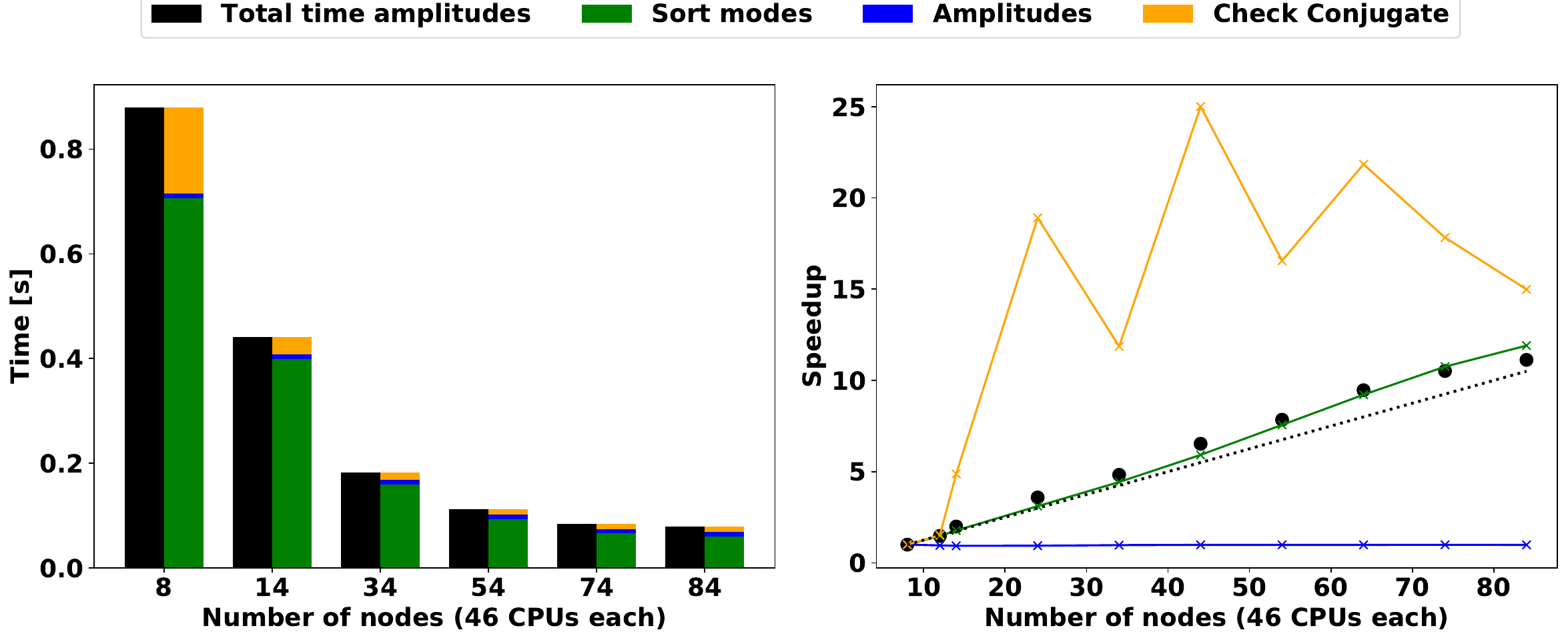}
    \caption{Profiling of the amplitude computation and sorting (left) and speedup of each of the functions involved (right)}
    \label{fig:speedup_amplitudes}
\end{figure}

\autoref{fig:speedup_amplitudes} shows that most of the time is used for sorting the modes according to the amplitude vector and checking that all subdomains have the imaginary part of the same sign when reordering conjugate pairs of modes. As the local size of the problem gets smaller, it takes less time to sort the modes out. This process is also affected by the the order of the modes before the sorting process, which is completely random. Thus, the cases with hyper-speedup are those in which more modes were already in their correct position before reordering. The actual computation of the amplitudes (\autoref{alg:jovanovic}) does not play any role in the performance of the DMD as it is done in serial by all ranks in a negligible amount of time.

%The sorting time and the randomness of the original order also dominate the performance of the amplitude timer in both weak scaling tests. The difference between them is that in the weak scaling regarding the number of rows, the time to sort the modes only depends on the original order of the modes and in the weak scaling regarding the number of columns the sorting time is reduced as the number of rows per processor gets smaller.

\subsection{Spectral proper orthogonal decomposition}
\begin{figure}
    \centering
    \includegraphics[width=0.95\textwidth]{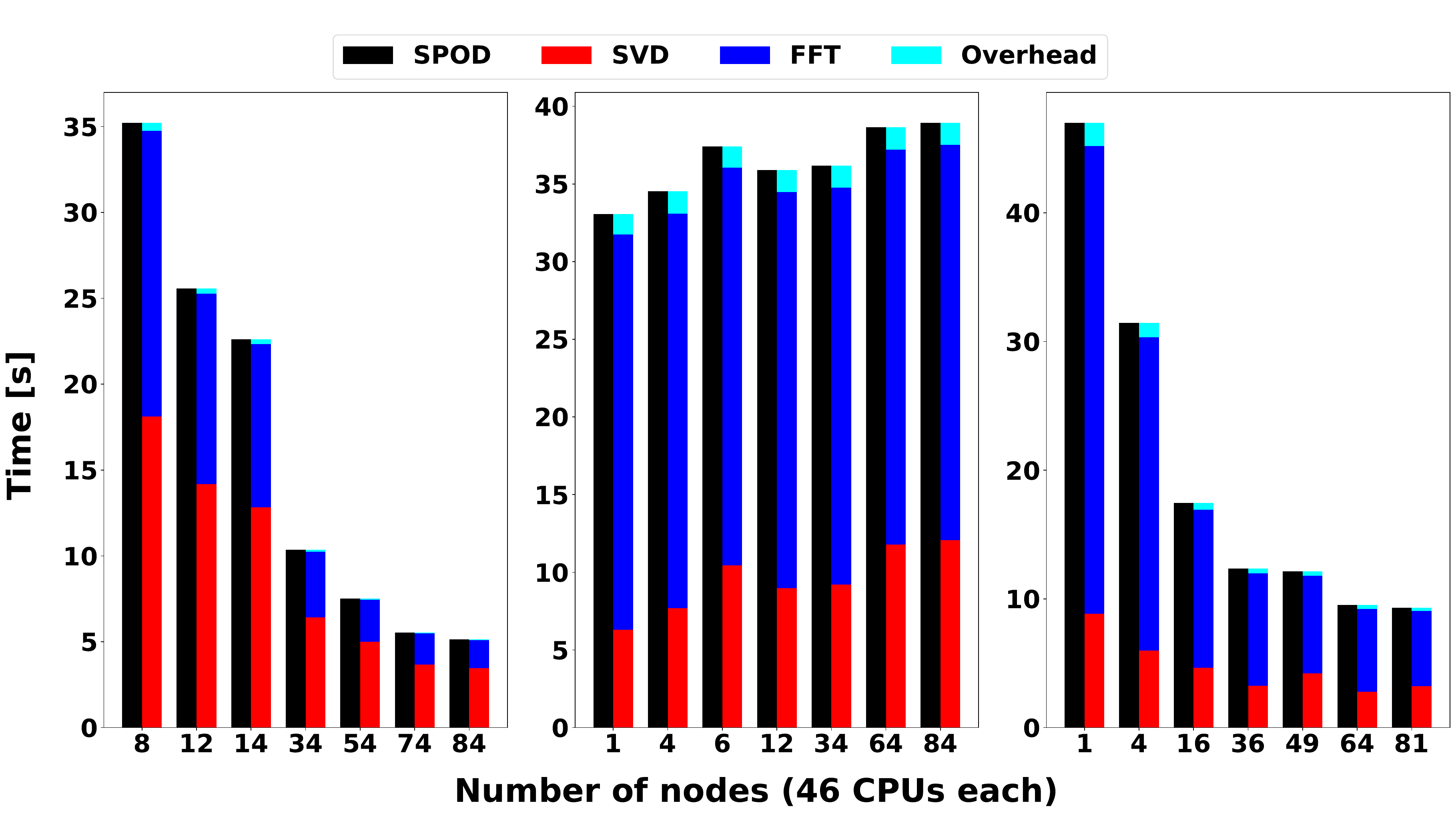}
    \caption{Profiling of the SPOD for the strong scaling (left), weak scaling concerning $M$ (middle) and weak scaling concerning $N$ (right) }
    \label{fig:profiling_SPOD}
    \vspace{0.1cm}
    \begin{tabularx}{\textwidth}{p{3.3cm}XXXXXXX}
    \hline
     & \textbf{8} & \textbf{12} & \textbf{14} & \textbf{34} & \textbf{54} & \textbf{74} & \textbf{84} \\
    \hline
    \textbf{SVD} & 51.4 & 55.4 & 56.8 & 61.9 & 66.6 & 66.3 & 67.5 \\
    \textbf{FFT} & 47.3 & 43.4 & 42.0 & 37.0 & 32.5 & 32.8 & 31.6 \\
    \textbf{Overhead} & 1.3 & 1.2 & 1.2 & 1.1 & 1.0 & 0.9 & 0.9 \\
    \hline
    \end{tabularx}    

    \captionof{table}{Percentage of execution time taken by each function for several numbers of nodes for the strong scaling in the SPOD}
    \label{tab:strong_SPOD}
    \vspace{0.1cm}
    \begin{tabularx}{\textwidth}{p{3.3cm}XXXXXXX}
\hline
 & \textbf{1} & \textbf{4} & \textbf{6} & \textbf{12} & \textbf{34} & \textbf{64} & \textbf{84} \\
\hline
\textbf{SVD} & 19.1 & 22.3 & 27.9 & 25.0 & 25.5 & 30.5 & 31.0 \\
\textbf{FFT} & 76.9 & 73.6 & 68.4 & 71.1 & 70.6 & 65.8 & 65.4 \\
\textbf{Overhead} & 4.0 & 4.1 & 3.7 & 3.9 & 3.9 & 3.7 & 3.6 \\
\hline
\end{tabularx}

    \captionof{table}{Percentage of execution time taken by each function for several numbers of nodes for the weak scaling concerning the number of rows in the SPOD}
    \label{tab:weak_SPOD}
    \vspace{0.1cm}
    \begin{tabularx}{\textwidth}{p{3.3cm}XXXXXXX}
\hline
 & \textbf{1} & \textbf{4} & \textbf{16} & \textbf{36} & \textbf{49} & \textbf{64} & \textbf{81} \\
\hline
\textbf{SVD} & 18.8 & 19.0 & 26.6 & 26.4 & 34.7 & 29.2 & 34.7 \\
\textbf{FFT} & 77.3 & 77.4 & 70.3 & 70.6 & 62.6 & 67.8 & 62.5 \\
\textbf{Overhead} & 3.9 & 3.6 & 3.1 & 3.0 & 2.7 & 3.0 & 2.8 \\
\hline
\end{tabularx}

    \captionof{table}{Percentage of execution time taken by each function for several numbers of nodes for the weak scaling concerning the number of columns in the SPOD}
    \label{tab:weak2_SPOD}
\end{figure}

\autoref{fig:profiling_SPOD} shows the performance of the SPOD implementation in the three scaling tests while tables \ref{tab:strong_SPOD}, \ref{tab:weak_SPOD} and \ref{tab:weak2_SPOD},el  present the time percentage of each function for every test. The FFT function is executed once in each point for all blocks, $M_i\times N_{Blks}$. The SVD is executed once per frequency over a matrix of size $\left(M,\text{} N_{Blks}\right)$. The overhead accounts for memory allocation, variable declarations and the division of the original array $\cal{D}$ in the different blocks needed for the SPOD computation. The overhead reduces in the strong and the weak concerning $N$ scaling tests because the size of the resulting arrays is also smaller.

\begin{figure}
    \centering
    \includegraphics[width=\textwidth]{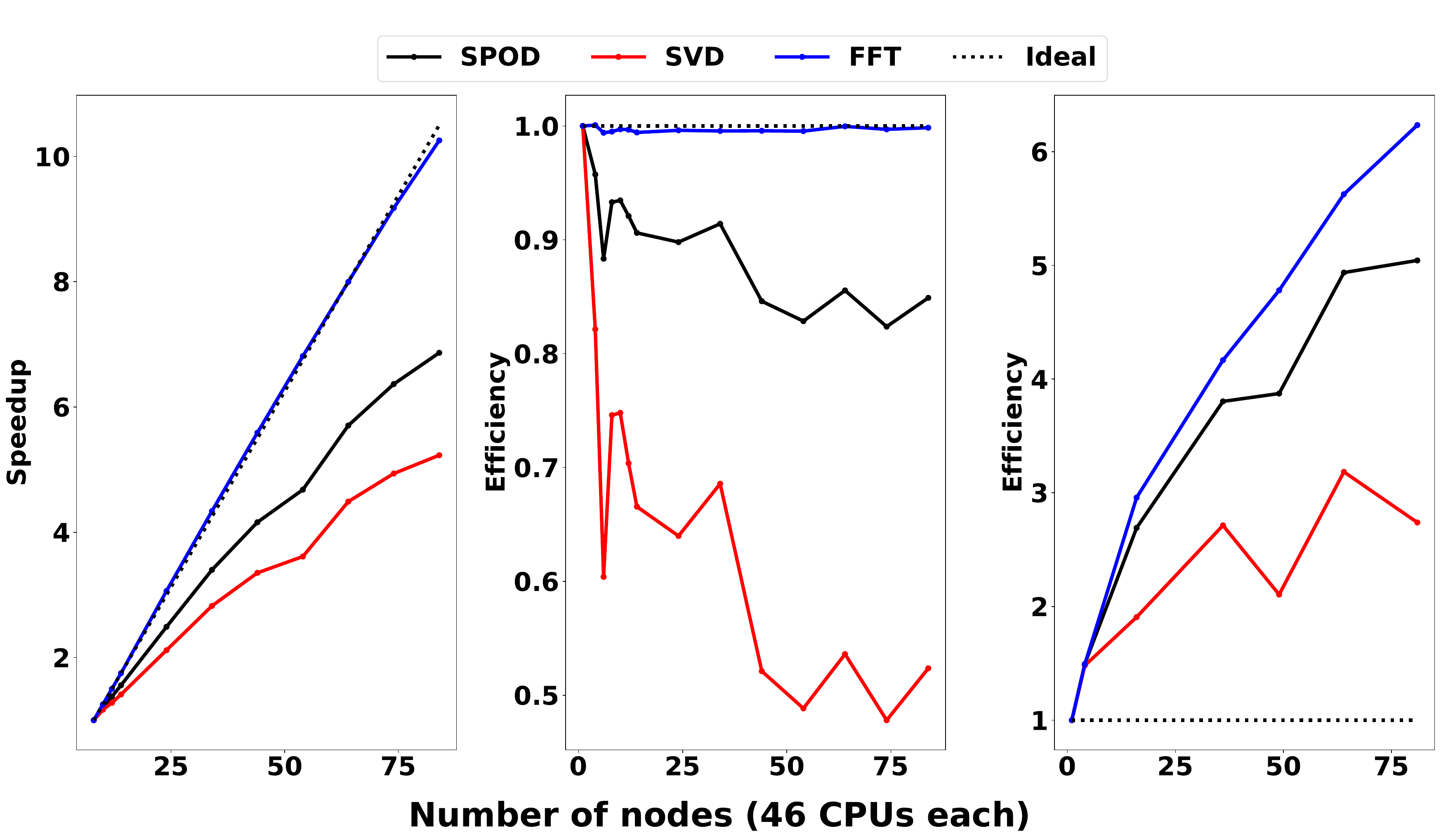}
    \caption{Speedup of the SPOD for the strong scaling (left) and efficiency for weak scaling concerning $M$ (middle) and weak scaling concerning $N$ (right) }
    \label{fig:speedup_SPOD}
\end{figure}

\autoref{fig:speedup_SPOD} presents the speedup for the strong scaling test in the SPOD and the efficiency for the two weak scaling analyses. The FFT has a perfect speedup and efficiency because the number of calls to the FFT per processor decreases with the number of processors, $P$. Other than that, the SVD speedup has the same limitations that have been discussed for the POD and DMD.

The huge efficiency for the case of the weak scaling concerning $N$ is justified by the fact that the dependence of the SVD on the number of columns is not purely quadratic (as discussed for the POD) and also because the load of the FFT diminishes drastically as the number of rows to transform to the frequency domain decreases.

\section{Conclusions}
\label{sec:concl}
%% Arnau & Benet, examples  + scalability
The present work introduces pyLOM \cite{pyLOM}, an open source python library which entails all the necessary tools to compute the proper orthogonal decomposition (POD), dynamic mode decompoisition (DMD) and spectral proper orthogonal decomposition (SPOD) in a high performance computing environment. 

%The objective of POD is finding an orthogonal basis of vectors that can represent the dataset with optimality, DMD aims to find a linear mapping that connects the snapshots as a Krylov sequence, and SPOD looks for the orthogonal vectors that optimally represent the second-order space-time statistics. All in all, the three algorithms need to decompose a matrix containing the snapshots from the data into a set of spatial correlations together with its dynamic information. These spatial correlations are found using the single value decomposition (SVD). 

%pyLOM is specifically focused for fluid dynamics applications. In those cases, each row of the snapshot matrix contains the data from a grid point while each column represents one snapshot. This data organization yields a tall and skinny matrix because the number of grid points, rows, is much larger than the number of instants, columns. This matrix is split row-wise so that each rank only needs to manage the information of a subset of grid points.

These three algorithms decompose a matrix containing the snapshots from the data into a set of spatial correlations together with its dynamic information through an SVD. To efficiently compute the SVD in parallel, the equivalence between the QR and SVD is used for tall and skinny matrices \cite{chan1982improved}, where the QR factorization is obtained through the algorithm presented Demmel et al. \cite{demmel_communication-optimal_2012}, and using a binary tree reduction and broadcasting algorithm.

The POD and DMD in pyLOM have already been used in previous studies \cite{eiximeno2022wake,richi,eiximeno2023hybrid,miro2023self}; this work is the first to introduce and validate the SPOD implementation of the library. Firstly,  the analysis of Begiashvili et al. \cite{begiashvili2023data} is reproduced to compare the modes and spectrum of the flow around a circular cylinder at Reynolds number $Re_D = DU_{\infty}/\nu=100$. Then, the SPOD algorithm is tested in a high performance computing environment to replicate the results obtained with POD and DMD of the flow around a circular cylinder at $Re_D = DU_{\infty}/\nu = 1\times 10 ^4$ by Eiximeno et al. \cite{eiximeno2023hybrid} and in the Stanford diffuser at $Re_h = hU_b/\nu = 1\times 10 ^4$ by Miró et al. \cite{miro2023self}. The latter case has $1.14\times 10^8$ points and 1808 snapshots and it is the biggest done with pyLOM up to the moment of writing this manuscript. Using 10368 CPUs, the POD, DMD and SPOD took 81.08, 947.89 and 20.95 seconds, respectively. 

Strong and weak scalability tests have been performed to profile the library in detail. The strong scalability tests show that, when the problem size is reduced with the number of processors, the part of the code that cannot be parallelized is smaller than the 10\% in the three algorithms. On the other hand, when the load per processor remains constant, the weak scalability tests reveal that less than 1\% of the algorithms are executed in serial. 

An in depth profiling analysis reveals that the bottleneck of the three algorithms are the communications during the SVD computation. In the case of the POD, it accounts for nearly all the computing time except for a small overhead which is due to variables definition and memory allocation.  The communications cost in POD increases with the logarithm of the number of processors, which is in agreement with what Demmel et al. \cite{demmel_communication-optimal_2012} reported for the QR factorization algorithm. The speedup and efficiency of the DMD are slightly higher as the relevance of the SVD decreases to around 60 \% of the computational time. Most of the remaining time is dominated by the perfect scalability of the mode computation and the hyperscalcability of the amplitude computation and mode sorting. Only the parallel matrix-matrix product used for the linear mapping computation has lower parallel performance than the SVD because the cost of its communications scales linearly with the number of processors. The SVD loses even more relevance for the SPOD, which is governed by the perfect scalability of the Fast Fourier Transform (FFT).

Finally, a weak scaling analysis regarding the number of columns has been performed to show the dependence of the algorithms on the number of snapshots of the dataset, $N$. This test has been set so that the number of columns increases as the square root of the number of processors, however, the results shows that the actual dependence of the QR factorization with the number of columns is $N^{1.37}$. For the DMD, the mode computation is the subroutine with the greatest dependence on the number of columns and dominates the load for the larger cases. In the case of the SPOD, the FFT has a lower dependence than the SVD, making the algorithm less sensitive to an increase of the number of snapshots to analyze.

%% Acknowledgements
\section*{Acknowledgements}
The research leading to this work has been partially funded by the European Project NextSim which has received funding from the European High-Performance Computing Joint Undertaking (JU) under grant agreement No 956104 and co-founded by the Spanish Agencia Estatal de Investigacion (AEI) under grant agreements PCI2021-121962 and PCI2021-121937. This work was partially financially supported by the Ministerio de Economía, Industria y Competitividad, Secretaría de Estado de Investigación, Desarrollo e Innovación, Spain (refs PID2020-116937RB-C21 and PID2020-116937RB-C22). B. Eiximeno’s work was funded by a contract from the Subprograma de Ayudas Predoctorales given by the Ministerio de Ciencia e Innovación (PRE2021-096927). Oriol Lehmkuhl has been partially supported by a Ramon y Cajal postdoctoral contract (Ref: RYC2018-025949-I). The authors acknowledge the support of Departament de Recerca i Universitats de la Generalitat de Catalunya to the Research Group Large-scale Computational Fluid Dynamics (Code: 2021 SGR 00902) and the Turbulence and Aerodynamics Research Group (Code: 2021 SGR 01051). We also acknowledge the Barcelona Supercomputing Center for awarding us access to the MareNostrum IV machine based in Barcelona, Spain.

\newpage
\appendix
\section{Parallel QR decomposition}
\label{sec:qr}
\begin{algorithm}[H]
    \caption{Tall and skinny QR decomposition \citep{demmel_communication-optimal_2012}.}
    \label{alg:tsqr_svd}
    \setstretch{1.35}
    \SetAlgoLined
    \KwIn{${\cal{D}}_i$\tcp*{Data matrix dispersed on each processor}}
    \KwOut{$Q_i$, $R$\tcp*{Local result of Q and global result of R}}
    %\end{algorithm}
    %\begin{algorithm}
    %\setstretch{1.35}
    %\SetAlgoLined
    \SetKwFunction{tsqrsvd}{tsqr\_svd}
    \SetKwFunction{svd}{svd}
    \SetKwFunction{Fqr}{qr}
    \SetKwFunction{nextPower}{nextPowerOf2}
    \SetKwFunction{Fnlog}{log2}
    \SetKwProg{Fn}{Function}{:}{}
    %\Fn{\tsqrsvd{${\cal{D}}_i$}}{
    %$Q_w$ = $I_n$\tcp*{Preallocate QW to the identity matrix of size n.}
    [$Q_{1i}$, R] = \Fqr{${\cal{D}}_i$}\tcp*{QR decomposition on each processor}
    nextPower = \Fnlog\nextPower{nprocs}\tcp*{Number of levels}
    ilevel = 0\tcp*{Level counter}
    \tcc{Loop to reduce all the information to Rank 0}
    \For{blevel = 1; blevel $<$ nextPower; blevel $<<=$ 1 }{
    C[0 : n, :] = R\tcp*{Store R in the upper part of the C matrix.}
    prank = myrank \textbf{XOR} blevel\tcp*{Processor to communicate}
    %\tcc{Processor which sends and receives information}
    \If{myproc \textbf{AND} blevel}{
    \If{prank $<$ nprocs}{
    Send R to prank
    }
    }
    \Else{
    \If{prank $<$ nprocs}{
    Receive R from prank\\
    C[n + 1 : end, :] = R\tcp*{Store R}
    [$Q_{2i}$, R] = \Fqr{C}\tcp*{QR of the C matrix}
    $Q_{2l}$[2n$\cdot$ilevel : 2n$\cdot$ilevel + 2n, :] = $Q_{2i}$\tcp*{Store $Q_{2i}$}
    }
    }
    ilevel ++
    }
    %}
    ilevel = nlevels - 1\tcp*{Level counter}
    \end{algorithm}
    \begin{algorithm}[H]
    \setstretch{1.35}
    \SetAlgoLined
    \setcounter{AlgoLine}{23}
    \tcc{Loop to broadcast the information from Rank 0}
    \For{blevel = 1 $<<$ (nlevels - 1); blevel $\geq$ 1; blevel $>>=$ 1}{
    mask = blevel - 1\\
    \tcc{Check if the processor has to send or receive information}
    \If{myrank \textbf{AND} mask == 0}{
    C = $Q_{2l}$[2n$\cdot$ilevel : 2n$\cdot$ilevel + 2n, :]\\
    $Q_{2i}$ = C$Q_w$\\
    prank = myrank \textbf{XOR} blevel\tcp*{Processor to communicate}
    %\tcc{Processor which sends and receives information}
    \If{myrank \textbf{AND} blevel}{
    \If{prank $<$ nprocs}{
    Receive C from prank\\
    R = C[0 : n, :]\\
    $Q_w$ = C[n + 1 : end, :]\\
    }
    }
    \Else{
    \If{prank $<$ nprocs}{
    C[0 : n, :] = R\\
    C[n + 1 : end, :] = $Q_{2i}$[n + 1 : end, :]\\
    $Q_w$ = $Q_{2i}$[1 : n, end]\\
    Send C to prank
    }
    }
    }
    ilevel$--$
    }
    $Q_i$ = $Q_{1i}Q_w$\tcp*{Final Q on each processor}
    \Return{$Q_i$, $R$}
\end{algorithm}

\section{Proper orthogonal decomposition algorithm}
    \label{sec:pod}
    \begin{algorithm}[H]
        \caption{Proper orthogonal decomposition.}
        \label{alg:pod}
        \setstretch{1.35}
        \SetAlgoLined
        \KwIn{${\cal{D}}_i$\tcp*{Data matrix dispersed on each processor}}
        \KwOut{$U_i$, S, V\tcp*{POD modes dispersed on each processor, singular values and right singular vectors (not dispersed)}}
        %\end{algorithm}
        %\begin{algorithm}
        %\setstretch{1.35}
        %\SetAlgoLined
        \SetKwFunction{tsqrsvd}{tsqr\_svd}
        \SetKwFunction{svd}{svd}
        \SetKwFunction{Fqr}{qr}
        \SetKwFunction{nextPower}{nextPowerOf2}
        \SetKwFunction{Fnlog}{log2}
        \If{removeMean}{
            ${\cal{D}}_{mean} = \frac{1}{n_t} \sum_{i_t=0}^{i_t=n_t}{{\cal{D}}_i[:,i_t]}$\\
            $Y_i = {\cal{D}}_i - {\cal{D}}_{mean}$ 
        }
        \Else{
            $Y_i = {\cal{D}}_i$ 
        }
        $U_i$, S, V = \tsqrsvd{$Y_i$}\tcp*{Implementation in algorithm 1}
        
\end{algorithm}

\section{Dynamic mode decomposition algorithms}
\label{sec:dmd}
\begin{algorithm}[H]
    \caption{Dynamic mode decomposition.}
    \label{alg:dmd}
    \setstretch{1.35}
    \SetAlgoLined
    \KwIn{${\cal{D}}_i$, $r$\tcp*{Data matrix dispersed on each processor and truncation residual for the SVD}}
    \KwOut{$\varphi_i$, $\mu_{Re}$, $\mu_{Im}$, $b$\tcp*{DMD spatial modes dispersed on each processor, real and imaginary part of the eigenvalues and mode amplitudes (not dispersed)}}
    %\end{algorithm}
    %\begin{algorithm}
    %\setstretch{1.35}
    %\SetAlgoLined
    \SetKwFunction{tsqrsvd}{tsqr\_svd}
    \SetKwFunction{svd}{svd}
    \SetKwFunction{Fqr}{qr}
    \SetKwFunction{truncate}{truncate}
    \SetKwFunction{matmulparal}{matmul\_paral}
    \SetKwFunction{nextPower}{nextPowerOf2}
    \SetKwFunction{eigen}{eigen}
    \SetKwFunction{vandermonde}{vandermonde}
    \SetKwFunction{cholesky}{cholesky}
    \SetKwFunction{Fnlog}{log2}
    \SetKwFunction{amplitudejovanovic}{amplitude\_jovanovic}
    \If{removeMean}{
        ${\cal{D}}_{mean} = \frac{1}{n_t} \sum_{i_t=0}^{i_t=n_t}{{\cal{D}}_i[:,i_t]}$\\
        $Y_i = {\cal{D}}_i - {\cal{D}}_{mean}$ 
    }
    \Else{
        $Y_i = {\cal{D}}_i$ 
    }
    $U_i$, S, V = \tsqrsvd{$Y_i$[:,:-1]}\tcp*{SVD of the first N-1 columns of $Y_i$}
    $U_i$, S, V = \truncate($U_i$, S, V, r)\tcp*{Truncate according to r}
    
    $aux = $ \matmulparal($U^{T}_i$, $ Y_i[:,1:]$)\tcp*{Parallel matrix-matrix product described in algorithm 4}
    $A = aux(S^{-1}V)^{T}$ \tcp*{$size(A) = N_r\times N_r$ and is the same matrix in all processors}
    $\mu_{Re}$, $\mu_{Im}$, $w$ = \eigen{$A_t$}\tcp*{Eigendecomposition with LAPACK dgeev}
    $\varphi_i = Y_i[:,1:]V^TS^{-1}w/\mu$\tcp*{Computation of the DMD modes}
    $b$ = \amplitudejovanovic{$\mu_{Re}$, $\mu_{Im}$, $w$}\tcp*{Algorithm 5}
    Order the modes and eigenvalues according to the amplitudes
    \end{algorithm}
\newpage
\label{sec:matmulparal}
\begin{algorithm}[H]
    \caption{Matrix-matrix parallel product.}
    \label{alg:matmul_parañ}
    \setstretch{1.35}
    \SetAlgoLined
    \KwIn{$A_i$, $B_i$\tcp*{Matrices dispersed on each processor}}
    \KwOut{$C$\tcp*{Reduced result of the product}}
    %\end{algorithm}
    %\begin{algorithm}
    %\setstretch{1.35}
    %\SetAlgoLined
    \SetKwFunction{tsqrsvd}{tsqr\_svd}
    \SetKwFunction{svd}{svd}
    \SetKwFunction{Fqr}{qr}
    \SetKwFunction{truncate}{truncate}
    \SetKwFunction{matmulparal}{matmul\_paral}
    \SetKwFunction{nextPower}{nextPowerOf2}
    \SetKwFunction{eigen}{eigen}
    \SetKwFunction{vandermonde}{vandermonde}
    \SetKwFunction{allreduce}{all\_reduce}
    \SetKwFunction{Fnlog}{log2}
    \SetKwProg{Fn}{Function}{:}{}
    \Fn{\matmulparal{$A_i$, $B_i$}}{
    $aux_i = A_iB_i$\tcp*{Matrix product in each processor}
    C = \allreduce($aux_i$, op = sum)\tcp*{Reduction of the result}
    }
    \Return{C}
    \end{algorithm}

    \begin{algorithm}[H]
        \caption{DMD modes amplitude as in Jovanovic et al. \cite{jovanovic2014sparsity}.}
        \label{alg:jovanovic}
        \setstretch{1.35}
        \SetAlgoLined
        \KwIn{$\mu_{Re}$, $\mu_{Im}$, $w$\tcp*{Eigenvalues and eigenvectors}}
        \KwOut{$b$\tcp*{Amplitude of the modes}}
        %\end{algorithm}
        %\begin{algorithm}
        %\setstretch{1.35}
        %\SetAlgoLined
        \SetKwFunction{tsqrsvd}{tsqr\_svd}
        \SetKwFunction{svd}{svd}
        \SetKwFunction{Fqr}{qr}
        \SetKwFunction{truncate}{truncate}
        \SetKwFunction{matmulparal}{matmul\_paral}
        \SetKwFunction{nextPower}{nextPowerOf2}
        \SetKwFunction{eigen}{eigen}
        \SetKwFunction{vandermonde}{vandermonde}
        \SetKwFunction{allreduce}{all\_reduce}
        \SetKwFunction{Fnlog}{log2}
        \SetKwFunction{amplitudejovanovic}{amplitude\_jovanovic}
        \SetKwProg{Fn}{Function}{:}{}
        \Fn{\amplitudejovanovic{$\mu_{Re}$, $\mu_{Im}$, $w$}}{
        $V_{and}$ = \vandermonde($\mu_{Re}$, $\mu_{Im}$, $N_r$, $N-1$) \tcp*{Computation of the Vandermonde matrix}
        $P = w^{TC}w*V_{and}V_{and}^{TC}$\\
        $P$ = \cholesky($P$)\tcp*{Cholesky factorization of P done with LAPACK zpotrf}
        $G = SV$\\
        $q = diag(V_{and}Gw)^C$\\
        $b = P^{{TC}^{-1}}P^{-1}q$\tcp*{Amplitudes computation}
        }
        \Return{$b$}
    \end{algorithm}

    \section{Spectral proper orthogonal decomposition algorithm}
    \label{sec:spod}
    \begin{algorithm}[H]
        \caption{Spectral proper orthogonal decomposition.}
        \label{alg:spod}
        \setstretch{1.35}
        \SetAlgoLined
        \KwIn{${\cal{D}}_i$, $dt$, $npwin$, $nolap$\tcp*{Data matrix dispersed on each processor, timestep between snapshots, number of snapshots per window, number of overlapped snapshots in each window}}
        \KwOut{$P$, $L$, $f$\tcp*{SPOD spatial modes dispersed on each processor, energy and frequency of the modes}}
        %\end{algorithm}
        %\begin{algorithm}
        %\setstretch{1.35}
        %\SetAlgoLined
        \SetKwFunction{tsqrsvd}{tsqr\_svd}
        \SetKwFunction{svd}{svd}
        \SetKwFunction{floor}{floor}
        \SetKwFunction{ceil}{ceil}
        \SetKwFunction{fft}{fft}
        \SetKwFunction{mean}{mean}

        $nBlks = \floor(\frac{N-nolap}{npwin-nolap})$\tcp*{Number of blocks}
        $window = 0.54-0.46\cos(2\pi \left[0:1:npwin\right]/(npwin-1))$\tcp*{Hamming window function to give weights to each snapshot of the window}

        \If{removeMean}{
            ${\cal{D}}_{mean} = \frac{1}{n_t} \sum_{i_t=0}^{i_t=n_t}{{\cal{D}}_i[:,i_t]}$\\
            $Y_i = {\cal{D}}_i - {\cal{D}}_{mean}$ 
        }
        \Else{
            $Y_i = {\cal{D}}_i$ 
        }

        \For{each block in $nBlks$ ($i_{blk}$)}{
            $i_0 = i_{blk} \cdot (npwin - nolap)$\;
            \For{each point ($i_p$)}{
                $X_f = Y_i[i_p, i_0:i_0+npwin] \cdot window$\tcp*{Select the window and multiply its snapshot by its weight}
                $q_k[i_p, :] = \frac{\fft(X_f)}{\mean(window)\cdot npwin}$\tcp*{Compute the FFT}
            }
            $q_k[:, 1:-1] \times= 2$\;
            $Q[:, i_{blk}] \leftarrow$ Reshape and store $q_k$ in column-wise order\;
        }
    \end{algorithm}
    \begin{algorithm}[H]
    \setstretch{1.35}
    \SetAlgoLined
    \setcounter{AlgoLine}{18}        
    \SetKwFunction{real}{real}
        $f = \frac{\ceil(npwin / 2) + 1}{dt \cdot npwin}$\tcp*{Set up the frequency axis}
        $M_i$ is the number of points per processor\;
        \For{each frequency ($i_{freq}$)}{
            $Q_f = Q[i_{freq} \cdot M_i : (i_{freq} + 1) \cdot M_i, :]\frac{1}{\sqrt{nBlks}}$\;
            $U, S, V = \tsqrsvd(Q_f)$\;
            $P_i[:, i_{freq}] = \real(U)$\;
            $L[i_{freq}, :] = \left| S^2 \right|$\;
        }
        Order $P_i$, $L$ and $f$ according to the first column of $L$\;

    \end{algorithm}

%\section{Sample Appendix Section}
%Lorem ipsum dolor sit amet, consectetur adipiscing elit, sed do eiusmod tempor section \ref{sec:sample1} incididunt ut labore et dolore magna aliqua. Ut enim ad minim veniam, quis nostrud exercitation ullamco laboris nisi ut aliquip ex ea commodo consequat. Duis aute irure dolor in reprehenderit in voluptate velit esse cillum dolore eu fugiat nulla pariatur. Excepteur sint occaecat cupidatat non proident, sunt in culpa qui officia deserunt mollit anim id est laborum.

\newpage
 \bibliographystyle{elsarticle-num} 
 \bibliography{cas-refs}





% \end{thebibliography}
\end{document}